\documentclass[a4paper,useAMS,usenatbib]{mn2e}

\usepackage{graphicx}
\usepackage{amstext}
\usepackage{amsmath}
\usepackage{amssymb}

\begin{document}

\title[How the merger of two WDs depends on the mass ratio]{How the merger of
  two white dwarfs depends on their mass ratio: orbital stability and
  detonations at contact} 

\author[M. Dan, S. Rosswog, J. Guillochon, E. Ramirez-Ruiz]{Marius
  Dan$^{1}$\thanks{E-mail: m.dan@jacobs-university.de}, Stephan Rosswog$^{1,2}$,
  James  
  Guillochon$^{2}$, Enrico Ramirez-Ruiz$^{2}$\\
$^{1}$School of Engineering and Science, Jacobs University
Bremen, Campus Ring 1, 28759 Bremen, Germany\\
$^{2}$TASC, Department of Astronomy and Astrophysics, University
of California, Santa Cruz, CA 95064}

\date{Accepted ?. Received ?; in original form ?}

\pagerange{\pageref{firstpage}--\pageref{lastpage}} \pubyear{2011}

\maketitle

\label{firstpage}

\def\msun{$M_{\odot}$}
\def\Msun{$M_{\odot}$ }

\begin{abstract}
Despite their unique astrophysical relevance, the outcome of 
white dwarf binary mergers has so far only been studied for a very restricted number of 
systems. Here we present the results of 
a   survey with more than two hundred simulations systematically scanning the white dwarf binary parameter space. We consider white dwarf masses ranging from 0.2 to 1.2 \Msun and 
account for their different chemical compositions. We find excellent agreement
with the orbital evolution predicted by mass transfer stability analysis. Much of our effort in this paper is dedicated to determining which binary systems are prone to a thermonuclear explosion just prior to merger or at surface contact. We find that a large fraction of He-accreting binary systems explode: all dynamically unstable systems with accretor 
masses below 1.1 \Msun and donor masses above $\sim$ 0.4 \Msun are found to trigger a helium 
detonation at surface contact. A substantial fraction of these systems could explode at  earlier times via detonations induced by instabilities in the accretion stream, as we have demonstrated in our previous work. We do not find  definitive evidence for an explosion prior to merger or at surface contact in any of the studied double carbon-oxygen systems. Although we cannot 
exclude their occurrence if some helium is present, the available parameter space for a successful detonation in a white dwarf binary  of pure carbon-oxygen composition is small.
We demonstrate that a wide variety  of dynamically unstable systems are viable type Ia candidates. The next decade thus holds enormous promise for the study of these events, in particular with the advent of wide-field synoptic surveys allowing a detailed characterization of their explosive properties.
\end{abstract}


\section{Introduction}
\label{sec:intro}

Type Ia supernovae are lynchpins in measuring distances on cosmological scales  
\citep{riess98,perlmutter98}. These events are bright enough to be seen
from a great distance such that many thousands have been catalogued by modern
transient surveys such as PTF \citep{rau09a}, CfA \citep{hicken09a,hicken09b},
and LOSS \citep{leaman11}. Despite the large number of detected events, the possible
progenitor(s) of type Ia supernovae are still not known \citep{howell11}.

A strong constraint comes from the type Ia event recently discovered in
M101, which placed an upper limit of 0.02 $R_\odot$ on the size of star
that exploded \citep{nugent11,bloom11}. This all but eliminates the  
possibility that the companion was a red giant \citep{li11}, and restricts the
pre-explosion mass transfer rate from any non-giant companion to a narrow range of values
\citep{horesh11,chomiuk12}. Observations indicate that the event is a fairly standard type Ia
\citep{brown11,bloom11}, and thus the  
particular progenitor channel that led to this event is likely  to be similar to a significant 
fraction of the type Ia events observed. The strong restrictions and/or elimination of 
non-degenerate companions compels us to explore the possibility that the companion 
is also a degenerate star. This double degenerate (DD) scenario for type Ia
would also be consistent with the observed delay time distribution, which is
approximately  
proportional to $t^{-1}$ \citep{totani08,maoz10,ruiter10}. This progenitor channel would further 
explain why hydrogen has never been observed in a type Ia \citep{howell11}.
Moreover, the recently observed super-Chandrasekhar mass events 
\citep{howell06,scalzo10,taubenberger11} give further credence to the idea that at 
least some fraction of type Ia may be attributed to the DD scenario.

Because white dwarfs (WDs) represent the final evolutionary state of the vast majority of 
stars, they are also the most common constituents of binaries containing compact stellar 
remnants. WDs are repositories of thermonuclear fuel and it is not surprising that there are
several pathways that can produce a powerful explosion. In fact, it seems that all but the 
lowest mass WDs ($< 0.2$ \msun) are capable of being ignited by impacts that are of order 
their own escape speeds \citep{rosswog09b,raskin09}. Moreover, the resulting theoretical lightcurves and spectra
show excellent agreement with observed type Ia. Unfortunately, these collision events are most likely
too rare to explain a substantial fraction of the observed type Ia events.

More frequently, a WD will find itself as a companion to another WD after both
stars have evolved  
off the main sequence or after a dynamical exchange in a dense stellar environment \citep{shara02,lee10}. Binary 
population modeling, which is calibrated by the currently observed systems, estimates around 
$10^8$ double WDs (DWDs) in the Milky Way \citep{nelemans01a}. The rates at which these systems 
come into contact, although of great interest for the formation of interacting binaries and 
possibly explosive mergers, remain uncertain, because they depend primarily on the distance 
of the two WDs after leaving the poorly understood common envelope phase
\citep{woods11}. Once the WDs have exited this phase,  
the two WDs are driven together via gravitational wave  radiation, the signature of which is 
expected to be a dominant component of the ambient gravitational wave sky 
\citep[e.g.][]{lipunov87,hils90,nelemans04,liu10}.

A critical stage in the evolution of a compact WD binary is the period just after mass transfer 
has been initiated. For the binary to survive, the mass transfer must be
stable, which depends sensitively  
on the internal structure of the donor star, the binary mass ratio, and the
angular momentum transport 
mechanisms \citep[e.g.][]{marsh04,gokhale07}. The torques applied by a disk that may form from
accreted material, and the tides raised on the accretor, are crucial for the
binary's orbital stability 
\citep{iben98,piro11}. If these mechanisms are unable to widen the orbit, the
system will merge quickly. 
The subsequent  evolution can produce a R Corona Borealis star \citep{webbink84,clayton07,longland11}, 
trigger an explosion as a type Ia
\citep{webbink84,iben84,piersanti03b,piersanti03a,yoon07}, or undergo an
accretion-induced collapse to a neutron star  
\citep{saio85,nomoto91,saio04,shen11}.

While theoretical advances detailing the possible outcomes of WD mergers  have
been made in the last decade \citep[see review by][]{postnov06}, simulations have only
probed the parameter space of DD mergers for a very small number of systems, with
the principle emphasis being on systems with a total mass greater than the Chandrasekhar mass. We have
previously stressed the vital role of the numerical initial conditions in determining the outcome of a
merger  \citep{dan11}, simulations with approximate initial conditions start with too little angular
momentum and predict a merger that occurs much more rapidly than what is realized in nature. Our previous
results indicated that many DD systems with a donor composed largely of helium (He) are expected to be ignited
by an interaction between the accretion stream and the torus of material acquired from the donor
\citep{guillochon10}. Here we greatly expand upon our initial study with 225
new simulations that fully explore the possible combinations of WD donors and
accretors in the mass range between 0.2 and 1.2 \msun.  
We demonstrate that our systematic study exhibits the orbital stability trends
expected for merging DD systems, thus enabling us to evaluate which systems are expected to produce
strong thermonuclear events in the lead up to or at the point of merger. We find that many systems with
He donors may produce explosions, but that systems with carbon-oxygen (CO) donors are incapable of exploding without
the inclusion of a significant He layer.

In Section \ref{sec:nummeth}, we recap our numerical approach and describe how our simulations
cover the parameter space of DD systems. In Section \ref{sec:orbstab} we give a brief overview of how dynamical
orbital stability is determined and compare these expectations to our simulation results. Section \ref{sec:detcontact}
discusses detonations that are either triggered by the accretion stream or at
the moment of merger for both He and CO mass-transferring donor stars. We
summarize our findings and relate our models with type Ia observations in Section
\ref{sec:discussion}.  For reference, we provide  
fitting formulae bounding the region of parameter space where detonations are expected in Appendix \ref{sec:fitting}.

\begin{figure}
  \begin{center}
    \includegraphics[height=2.8in]{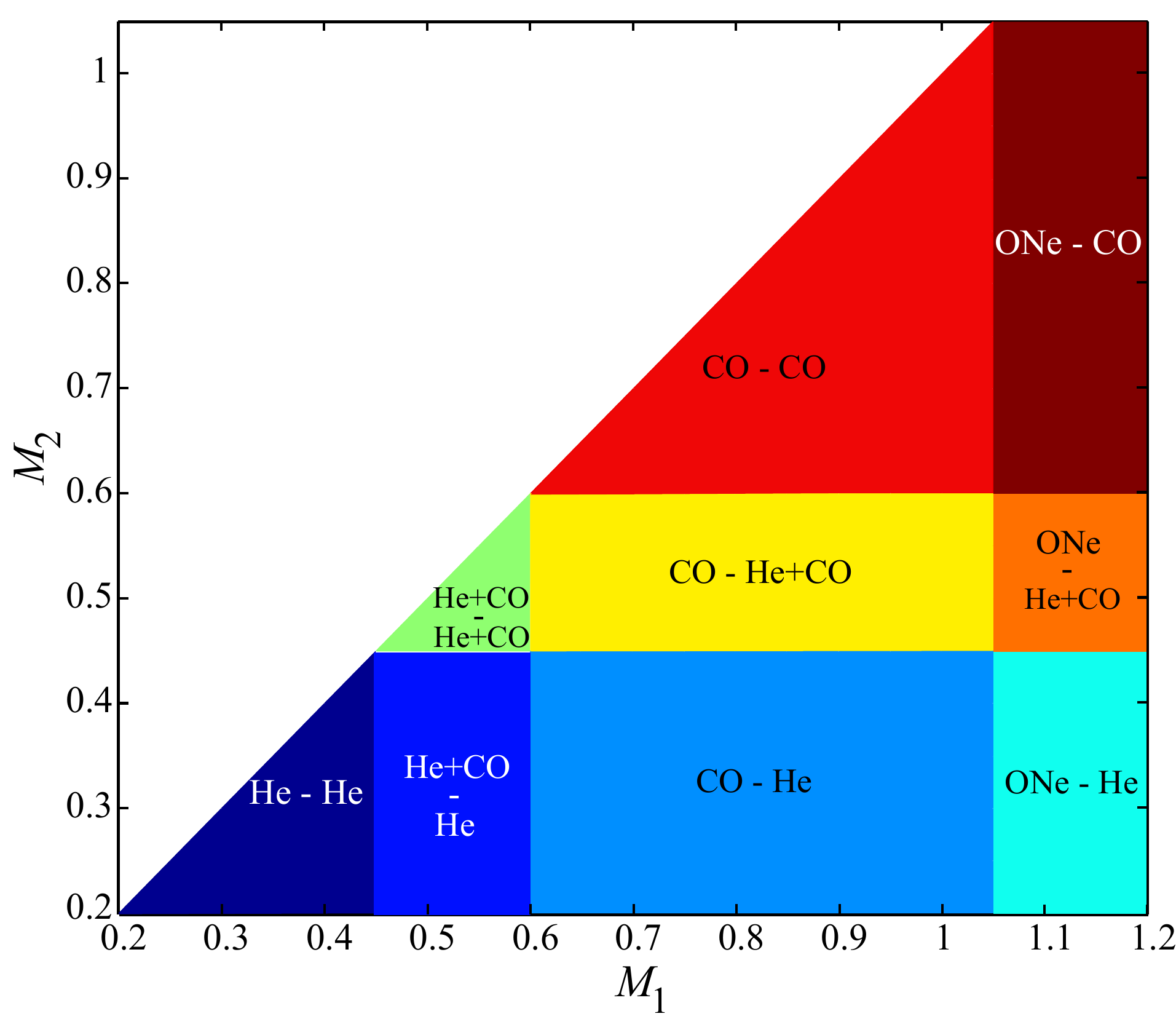}
  \end{center}
  \caption{The initial chemical composition of double WD systems with donor
    masses $M_2$ from 0.2 to 1.05 \Msun and accretor masses $M_1$ between 0.2
    and 1.2 \msun. See text for additional information.} 
  \label{fig:comp}
\end{figure}

\begin{figure*}
  \centerline{
      \includegraphics[height=1.9in]{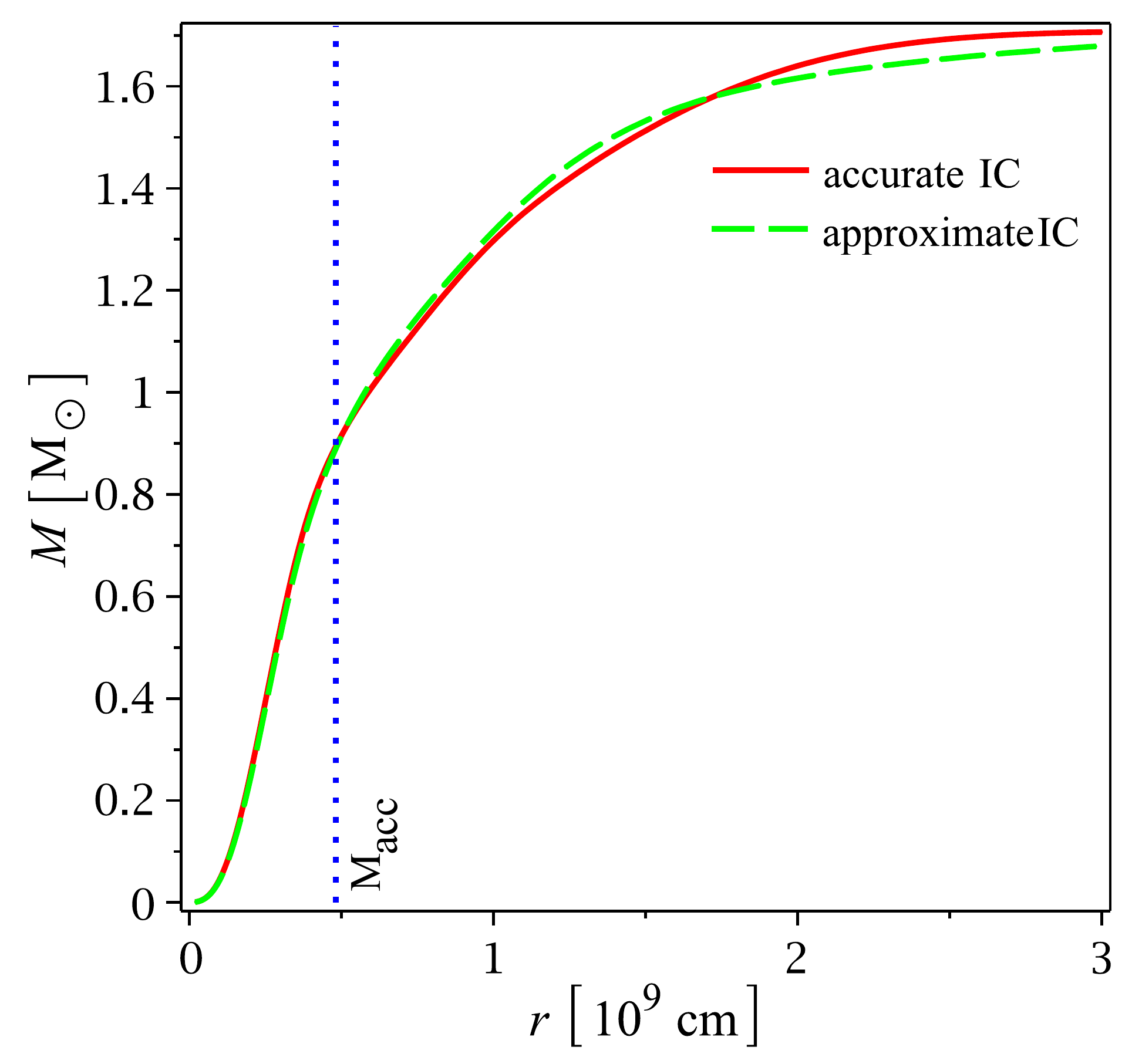}\hspace{0.5cm}
      \includegraphics[height=1.9in]{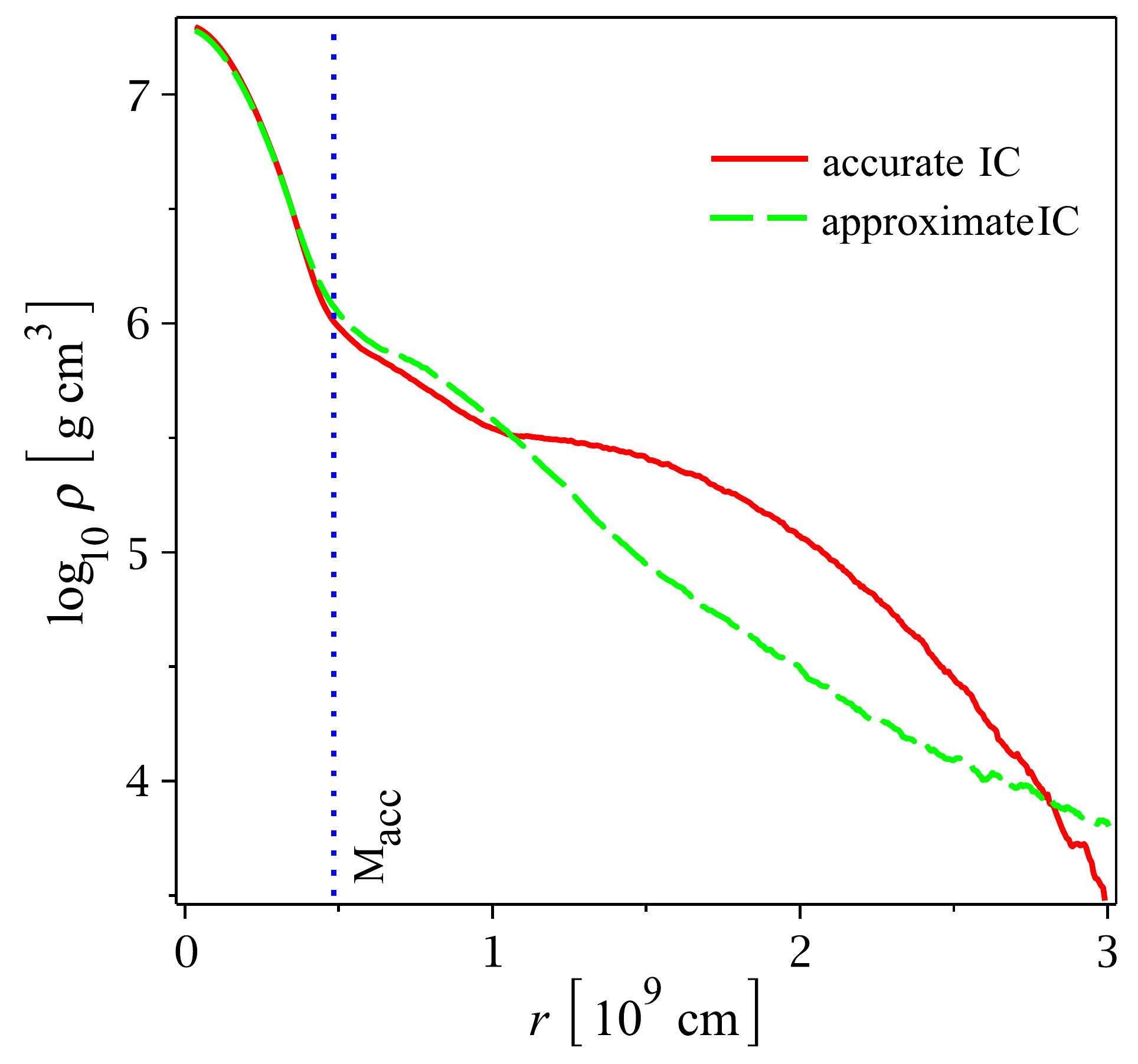}\hspace{0.5cm}
      \includegraphics[height=1.9in]{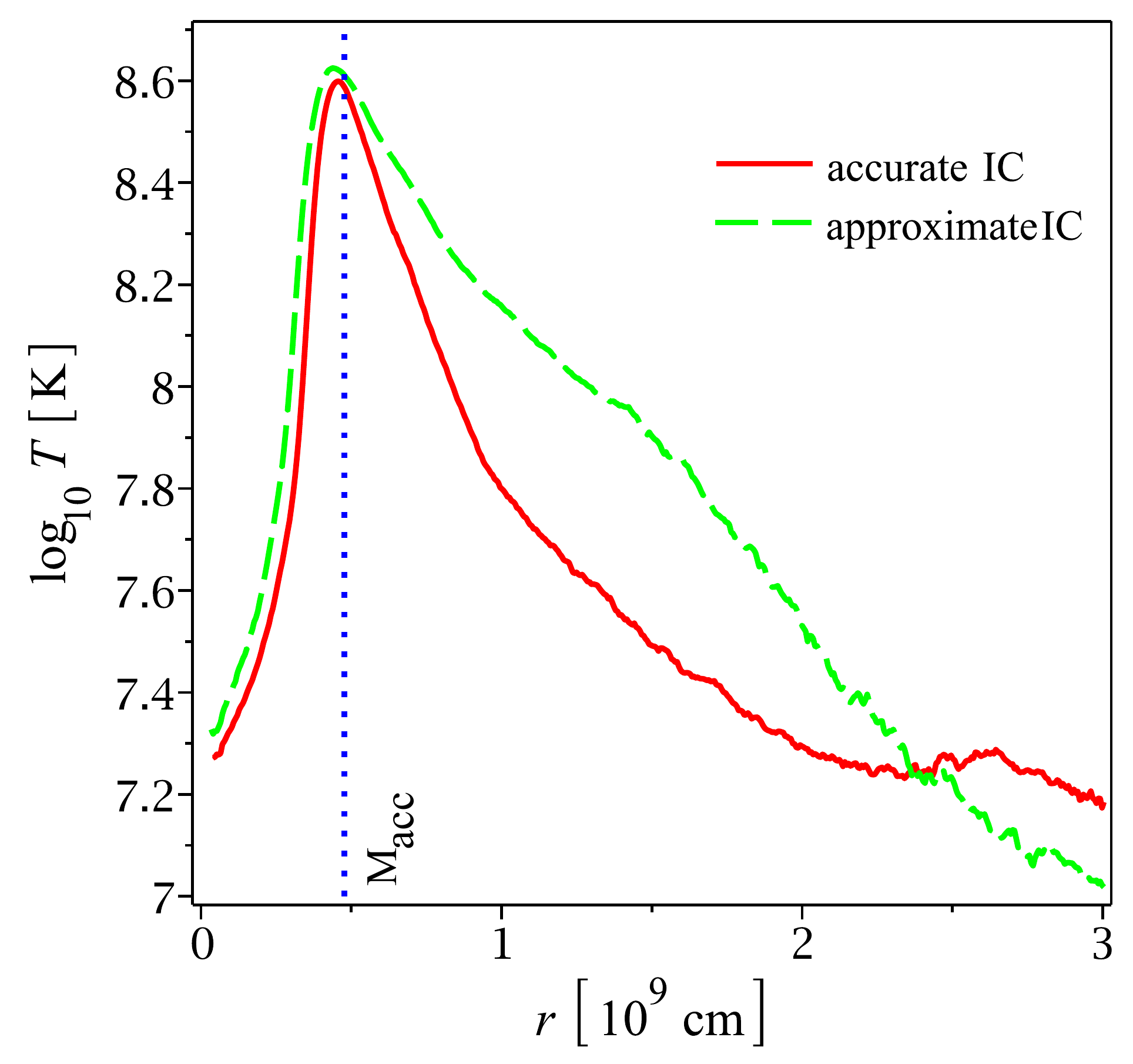}
  }
  \caption{Remnant profiles both with accurate (solid-red line) and
    approximate (dashed-green line) initial conditions (ICs) for the $0.81+0.9$
    \msun. Shown are 
    the spherically enclosed mass (left) and spherically averaged density
    (center) and temperature (right) vs. radius from the center of the accretor.}  
  \label{fig:compICs081}
\end{figure*}

\section{Numerical approach}
\label{sec:nummeth}

The simulations of this investigation are performed with a 3D smoothed
particle hydrodynamics (SPH) code \citep{rosswog08}, for recent 
reviews of the SPH method see e.g. \cite{monaghan05} and
\cite{rosswog09}. The orbital dynamics of a binary system is very sensitive to
the redistribution of angular momentum during mass transfer and 
therefore accurate numerical conservation is key to a reliable
simulation. SPH is a very powerful tool for this type of study because
the equations conserve angular momentum  by
construction, see e.g. Section 2.4 in \citet{rosswog09}. In practice,
the quality of conservation is determined by the force evaluation
\citep[in our case by the opening criterion of our binary tree][]{benz90}
and by the time stepping accuracy, both of which can be made arbitrarily accurate. Our
code uses an artificial viscosity scheme \citep{morris97} that reduces the
dissipation terms to a very low level away from discontinuities, together with
a switch \citep{balsara95} to suppress the spurious viscous forces in pure
shear flows. The system of fluid equations is closed by the Helmholtz equation
of state \citep{timmes00}. It accepts an externally calculated nuclear
composition and allows a convenient coupling to a nuclear reaction network.
We use a minimal nuclear reaction network \citep{hix98} to determine the evolution 
of the nuclear composition and to include the energetic feedback onto the gas
from the nuclear reactions. A set of only seven abundance groups greatly
reduces the computational burden, but still reproduces the energy generation
of all burning stages from He burning to NSE accurately. We use a binary tree
\citep{benz90} to search for the neighbor particles and to calculate the
gravitational forces. More details about our SPH code can be found in
\cite{rosswog08} and the provided links to the literature.

\subsection{Initial conditions}
\label{sec:ICs}

\begin{figure*}
  \begin{center}
    \begin{center}
      \begin{tabular}{cc}
        \includegraphics[height=2in]{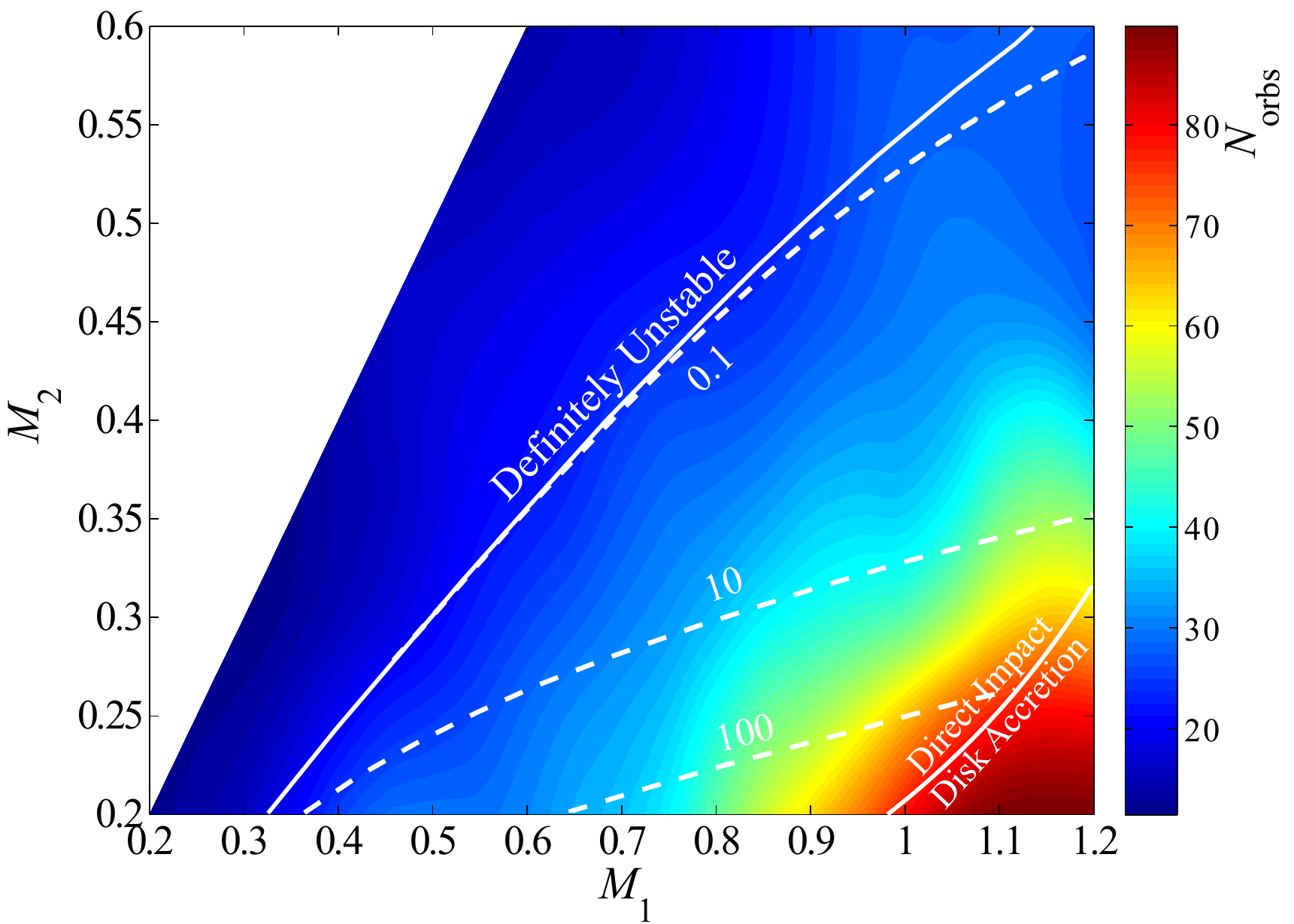}& 
        \includegraphics[height=2in]{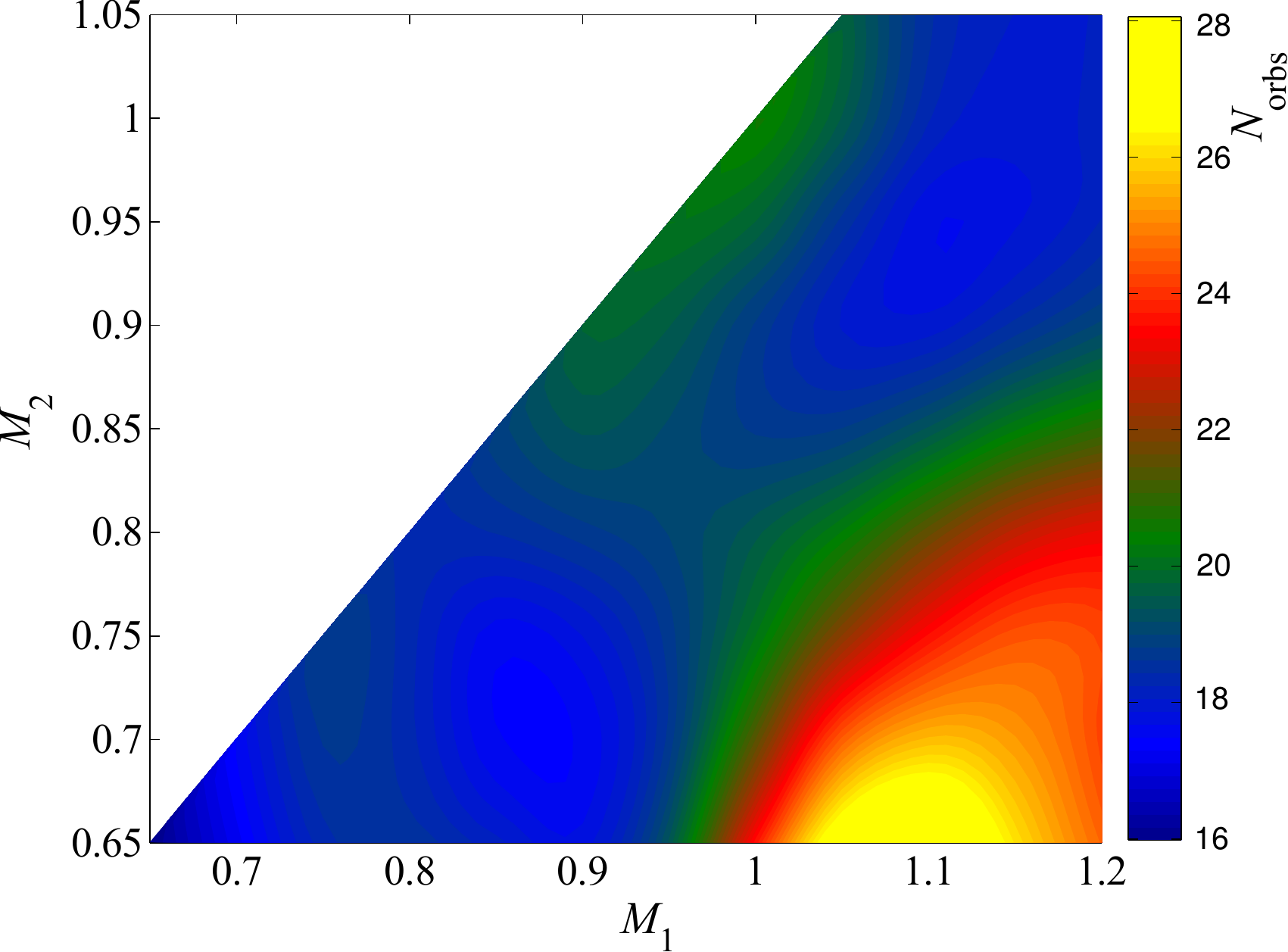}
      \end{tabular}
    \end{center}
  \end{center}
  \caption{Number of orbits of mass transfer for the He (left) and CO
    (right) donors. For He mass-transferring systems, the mass transfer
    stability limits are over-plotted. The region below the continuous lower line
    indicates where a disk forms, above this line the accretion stream
    directly impacts on the accretor. The mass transfer is guaranteed
    to be unstable  above the upper continuous line \citep{marsh04}. Dashed
    lines represent stability limits assuming a synchronization timescale of
    the accretor with the orbit of $0.1$, $10$ and $100\ {\rm yr}$.
    All systems with CO donors (right) belong to the unstable, direct impact
    regime of mass transfer.
  }
  \label{fig:norbs}
\end{figure*}

The WD donor mass distribution in our parameter space starts at 0.2 \Msun
and reaches up to 1.05 \Msun in steps of 0.05 \msun. For accreting WDs we
investigate those with masses between 0.2 and 1.2 \msun, again in steps of 
0.05 \msun. Accurate initial conditions (ICs) were constructed following the
procedure described in \cite{dan11}, here we only briefly summarize it.

For all our simulations the stars are initially synchronized, cold and
isothermal ($T=10^5$ K). In a first step, the stars are individually driven to
an accurate numerical equilibrium by adding a velocity-dependent damping term
to the momentum equation. Subsequently, we place the stars in the corotating
frame at an initial separation that is large enough to avoid any immediate
mass transfer. The orbital separation is then adiabatically reduced in a way
that the timescale over which the separation changes is much longer than the
dynamical timescale of the donor. During this process, spurious
oscillations are minimized by applying a damping force.  
When the first SPH particle reaches the inner Lagrangian point $L_1$, we stop
the relaxation process, transform the velocities to the fixed frame and set
this moment as the time origin ($t=0$) of the simulation. 
Most of the previous SPH WD merger calculations
\citep{benz90,segretain97,guerrero04,yoon07,aguilar09,pakmor10,pakmor11} 
started from rather approximate ICs. In \cite{dan11} we have shown in detail
that the initial conditions have a strong impact on all major aspects of the
simulations. 
Here, we only briefly illustrate the differences between the approximate and
our, accurate ICs in the remnant structures resulting from the merger of a
0.81 + 0.9 \Msun system in Figure~\ref{fig:compICs081}. When using the
accurate IC there is less mass in the disk and more mass in the trailing arm
than when starting from the approximate IC. The approximate IC also
overestimates the density and temperature in the region surrounding the
accretor. As we explain in Section~\ref{sec:discussion}, the pollution of the
ambient environment prior to any supernova-like event may have important
observational consequences.

The compositions used in this study are shown in Figure~\ref{fig:comp}.
Below 0.45 \Msun the stars are made of pure He. In principle, WDs with masses
below 0.45 \Msun could harbor heavier elements in their cores
\citep{iben85,han00,moroni09}, but 
\cite{nelemans01a} have shown that the probability for hybrid WDs in this
range is $4-5$ times smaller than for He WDs.
Based on \cite{rappaport09}, who found
that a 0.475 \Msun WD has a He envelope of about 30\% of its total mass,
between 0.45 and 0.6 \Msun we adopt a hybrid He-CO composition with a 
$\sim 0.1$ \Msun He mantle and pure CO core.
WDs between 0.6 and 1.05 \Msun are assumed to be made entirely of CO. According with \cite{iben85},
there is only a small amount of He at the surface of the CO core for this range of 
masses. It has been shown that tiny amounts of He can be important for
detonations of CO matter \citep[e.g.][]{seitenzahl09}, but for the
numerical resolution that we can afford in this large parameter study (20,000 SPH
particles/star) we cannot properly resolve the outer He layers.
For this range of masses there should be more Oxygen than Carbon
\citep{iben85}, and we have chosen mass fractions $X({}^{12}{\rm C})=0.4$ and  
$X({}^{16}{\rm O})=0.6$ uniformly distributed through the star, similar to
\cite{aguilar09}. 
For WDs with masses above 1.05 \Msun we use  
$X({}^{16}{\rm O})= 
0.60$, $X({}^{20}{\rm Ne})= 0.35$ and $X({}^{24}{\rm Mg})= 0.05$ (similar to the
composition found by \cite{gilpons01} for a 1.1 \Msun WD),
distributed uniformly throughout the entire star.

\section{Orbital stability}
\label{sec:orbstab}

After one or more common envelope phases, a white dwarf binary can emerge close enough for gravitational radiation to drive the binary towards shorter orbital periods until mass transfer is initiated. This occurs at orbital periods between 2 and 15 minutes depending on the characteristics of the donor. 
For the binary to survive, the mass transfer must be stable, which depends sensitively on  the binary mass ratio and 
the size of the  donor star. Dynamical instability is guaranteed for systems in which the mass ratio of the donor star to the accretor star $q \equiv M_2/M_1$ exceeds $2/3$ \citep[see e.g.][]{nelemans01b}. For $q<2/3$, the system can still survive as an interacting binary if orbital angular  momentum losses are balanced by other mechanisms. In many  instances, the formation of a  tidally truncated accretion disk  can efficiently return the advected angular momentum to the orbit.  The system can then survive and  evolve to longer periods as a result of mass transfer. 
On the other hand, mass transfer can proceed directly into the surface of the accretor. As a result, the system is destabilized by transforming orbital angular momentum into the accretor's spin.
A direct impact will occur if the radius of closest approach of the accretion stream $r_{\rm min}$ is less than the radius of the accretor $R_1$.
An estimate of $r_{\rm min}$ is given by \citep{lubow75,nelemans01b}
\begin{eqnarray}
\frac{r_{\rm min}}{a} &=&  0.04948 - 0.03815 \log q\nonumber\\
&\, & + 0.0475 \log^2 q - 0.006973\log^3 q\label{eq:rmin},
\end{eqnarray}
where $a$ is the orbital separation. The  two very-short period binaries HM Cnc and V407 Vul  belong to this so called    {\it direct impact} category as they both satisfy $r_{\rm min} < R_1$ \citep{marsh02,ramsay02}.

Tidal coupling can also return angular momentum to the orbit, heating both WD members  in the process. This tidal heating can influence the donor's mass transfer  response and, as a result,  affect the stability of the system. 
For the coupling to effectively stabilize the system, the synchronization torques must act on a timescale shorter than
the timescale over which the rate of mass transfer exponentiates.  Some WD binaries clearly avoid merger, as we see their survivors in the form of AM Canum Venaticorum (AM CVn) binaries \citep[see reviews by][]{warner95,nelemans05,solheim10}. Survival is however not guaranteed even when mass transfer is expected to be stable. Mass transfer can, for example,  exceed  the Eddington rate  or spark a thermonuclear explosion that might destroy the binary.

\begin{figure*}
  \centerline{
    \includegraphics[height=2.5in]{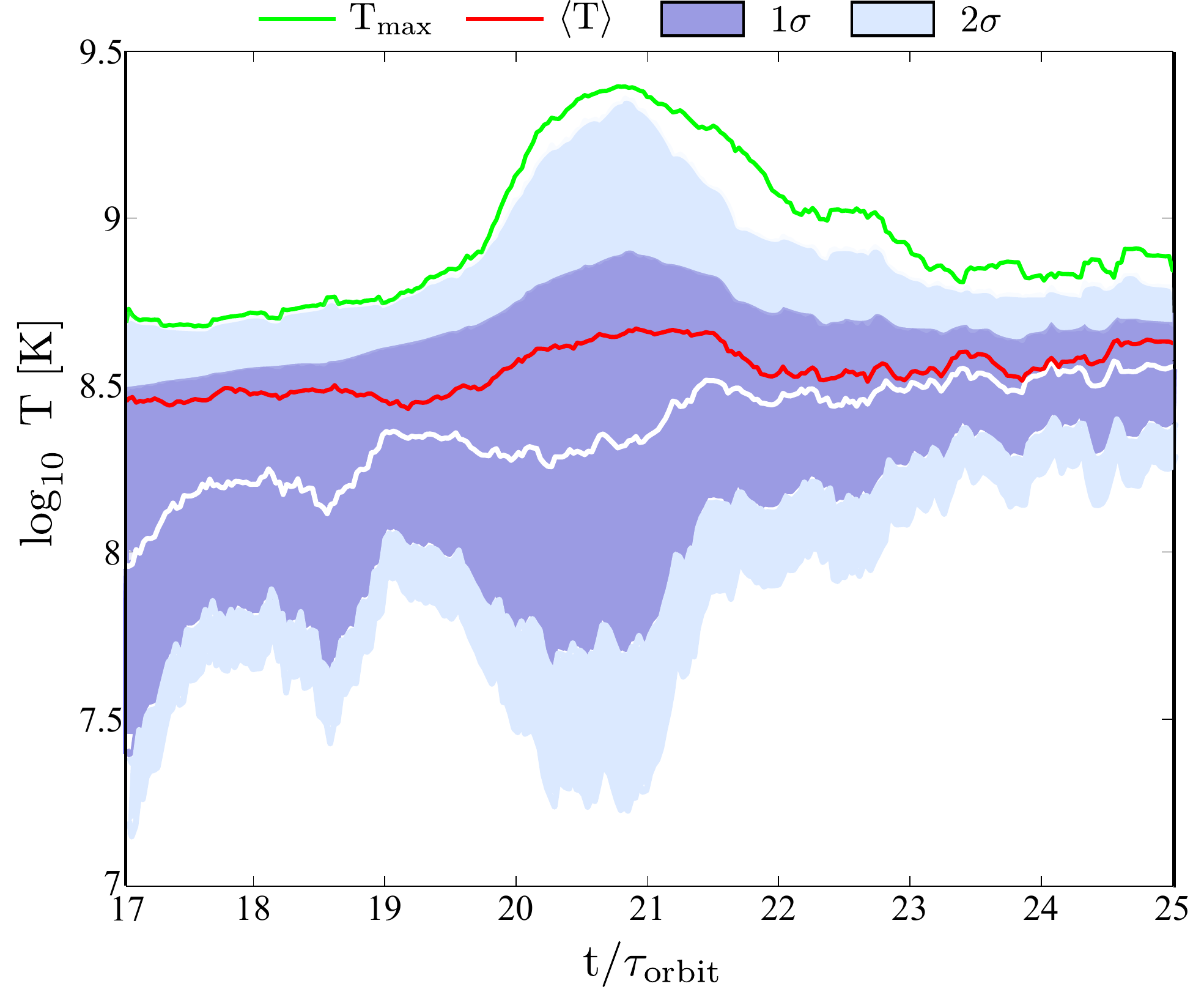}\hspace{0.5cm}
    \includegraphics[height=2.55in]{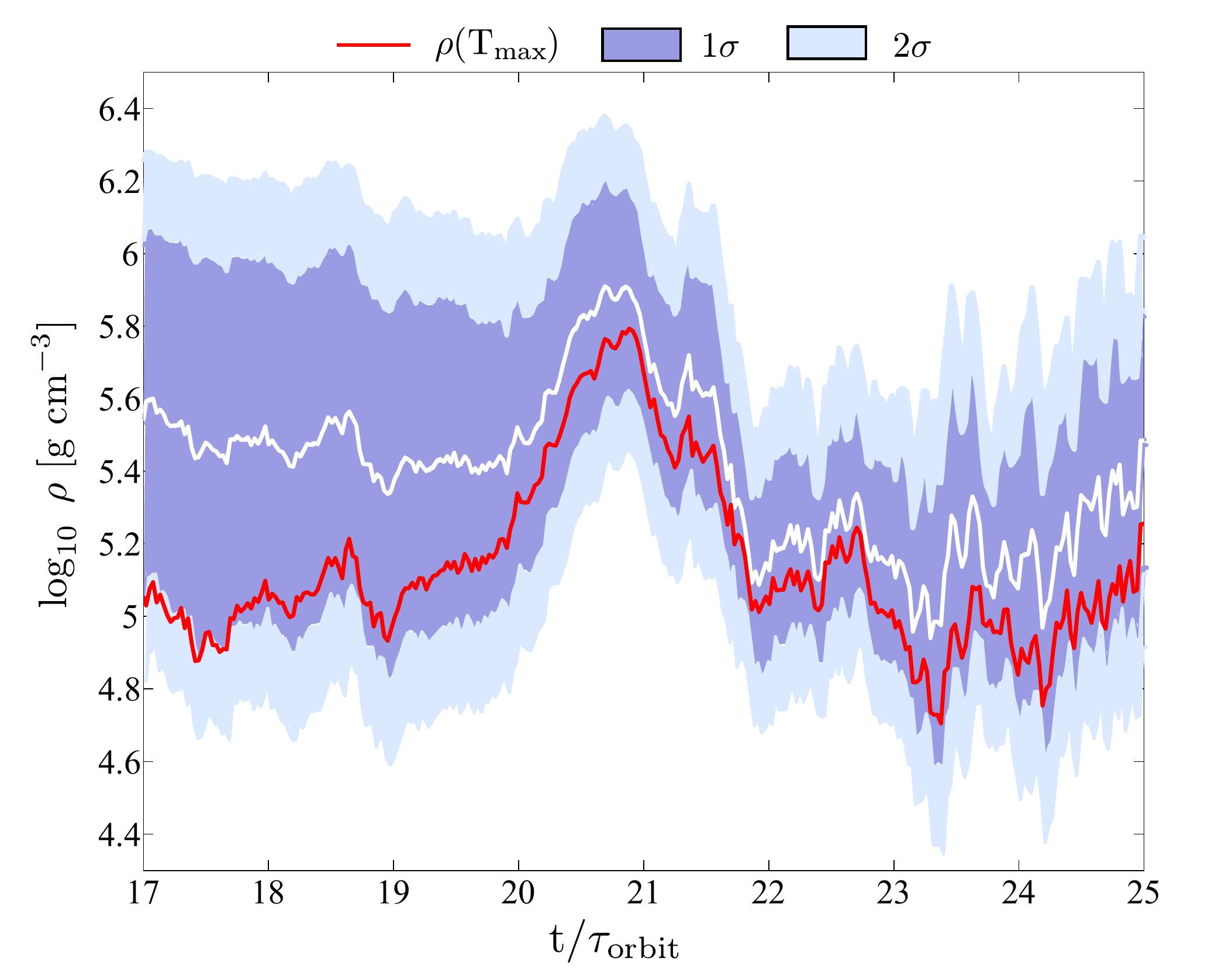}
  }
  \caption{
    Temperature (left) and density (right) evolution over the last 8
    orbital timescales for the $0.6+0.9$ \Msun system.  $T_{\rm max}$ is the
    maximum temperature, $\rho(T_{\rm max})$ its corresponding density and
    $\langle T\rangle$ is the SPH-smoothed temperature, see Eq. (\ref{eq:tsmooth}).
    $\langle T\rangle$ is computed using the particles within the smoothing
    length of particle with $T_{\rm max}$. Colored bands show $\pm$ 1, 2
    $\sigma$ (standard deviations) distributions around the mean values, the
    white line indicates the arithmetic temperature mean of the neighbors of
    the hottest particle.  
    Time is measured in units of the initial orbital period, $\tau_{\rm orbit}$. 
  }
  \label{fig:diffTrho}
\end{figure*}

In Figure~\ref{fig:norbs} we show the number of orbits each system survives  prior to merger. The mass transfer rates we are able to numerically resolve are so far  above the Eddington 
limit that photons are effectively trapped and, as a result, no significant mass loss is anticipated to occur. While all systems in our survey merge,  we expect that binaries  belonging to the disk accretion regime would  survive if we were able to resolve the  initial rate of mass transfer. In our  simulations, the degree to which the donor star overfills its Roche lobe is artificially enhanced by our finite resolution. This  results in a mass transfer rate that increases too quickly and produces  an under-massive  accretion disk that cannot  return sufficient angular momentum to the orbit to avoid merger.

Additionally, dynamically unstable systems are expected to be desynchronized at the initial
separations where numerically resolvable mass transfer sets in \citep{new97}. Constructing such initial conditions  is challenging as desynchronicity is a strongly destabilizing  effect.  While all our simulations were started with synchronized stars, the accretor becomes desynchronized after only a few orbital revolutions, and thus the systems quickly evolve to a state that is close to what is 
realized in nature. Despite the quick build-up to this desynchronized state, our simulations are able to capture the basic angular momentum transport mechanisms correctly as illustrated by the dramatic  increase in the number of orbits as we approach the disk accretion boundary.
In comparison with previous work at higher resolution in which the white dwarf binaries merge within only a few orbits, our results illustrate the importance of  carefully constructed initial conditions. 

\section{Detonations at contact}
\label{sec:detcontact}

In this section we investigate which binary systems could produce
immediate detonations at contact, first during the mass transfer, which in the
following we refer to as ``stream-induced detonations'', and, if detonations
from the stream interaction are avoided, during the final coalescence that we
call ``contact detonations''. 
In Sec.~\ref{sec:detcrit} we briefly summarize the detonation
criteria that we use in Sects.~\ref{sec:HeMT} and \ref{sec:COMT}
to discuss the detonation prospects for both He- and 
CO-transferring systems.

\begin{figure*}
  \centerline{
    \includegraphics[height=1.8in]{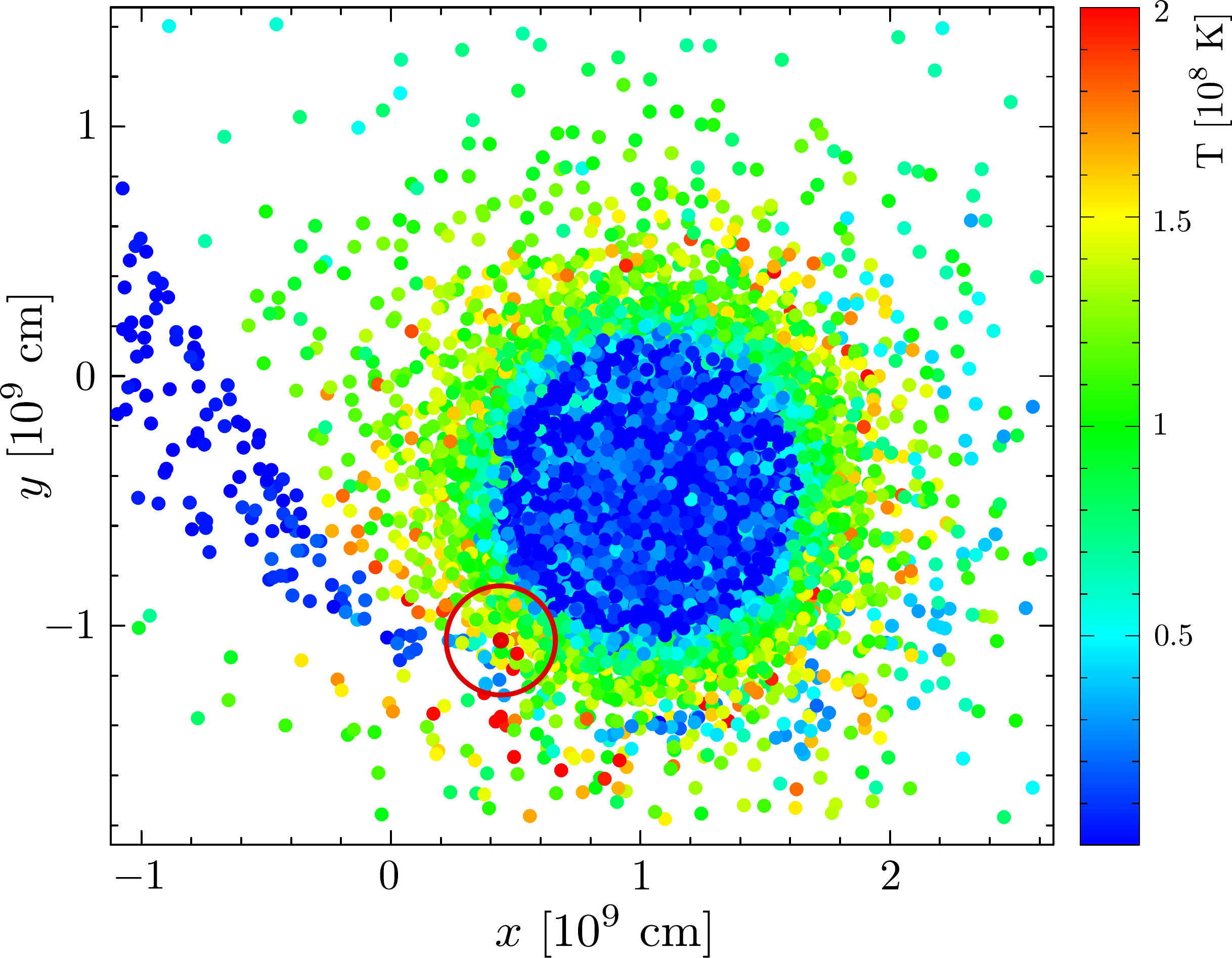}\hspace{0.15cm}
    \includegraphics[height=1.8in]{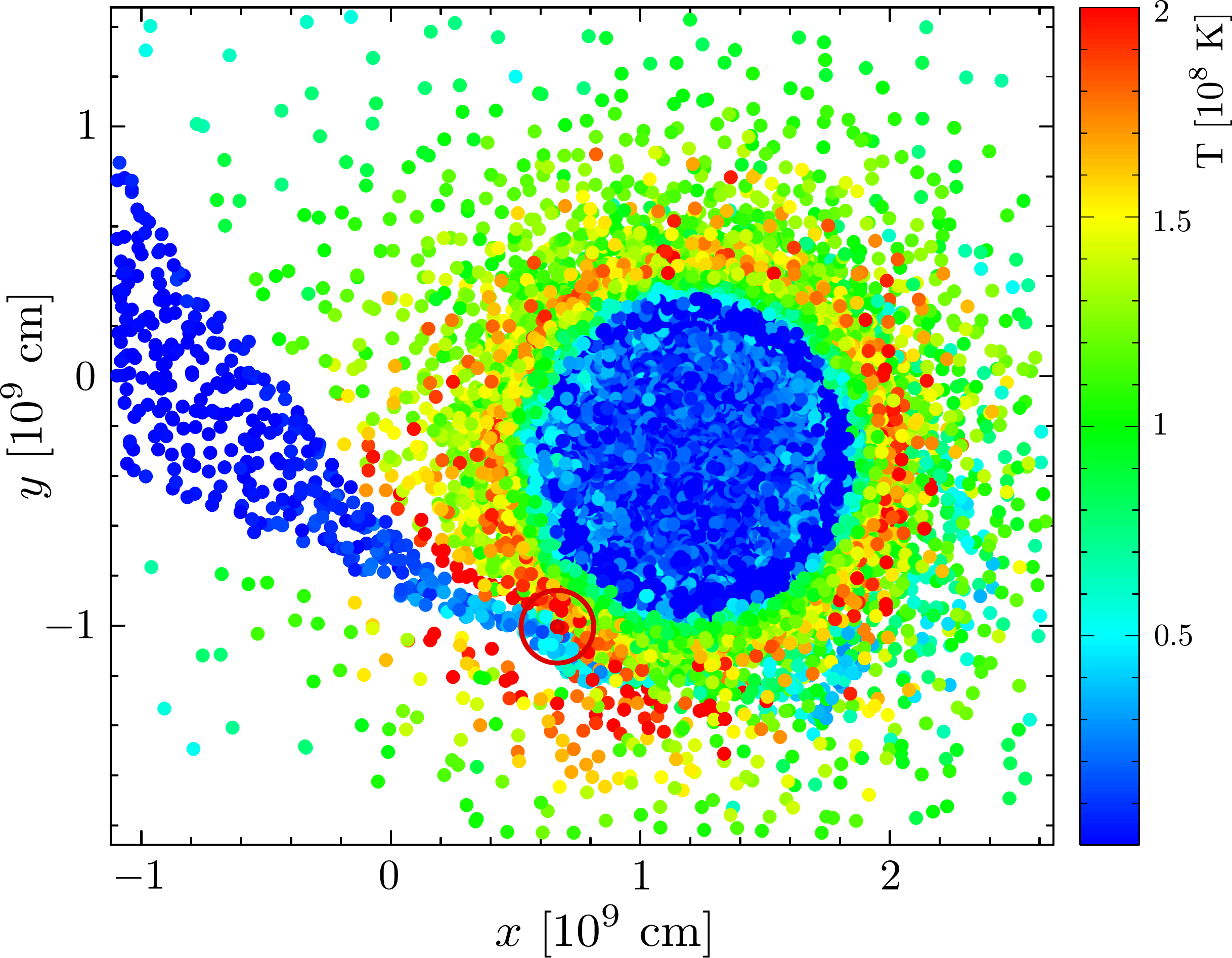}\hspace{0.15cm}
    \includegraphics[height=1.9in]{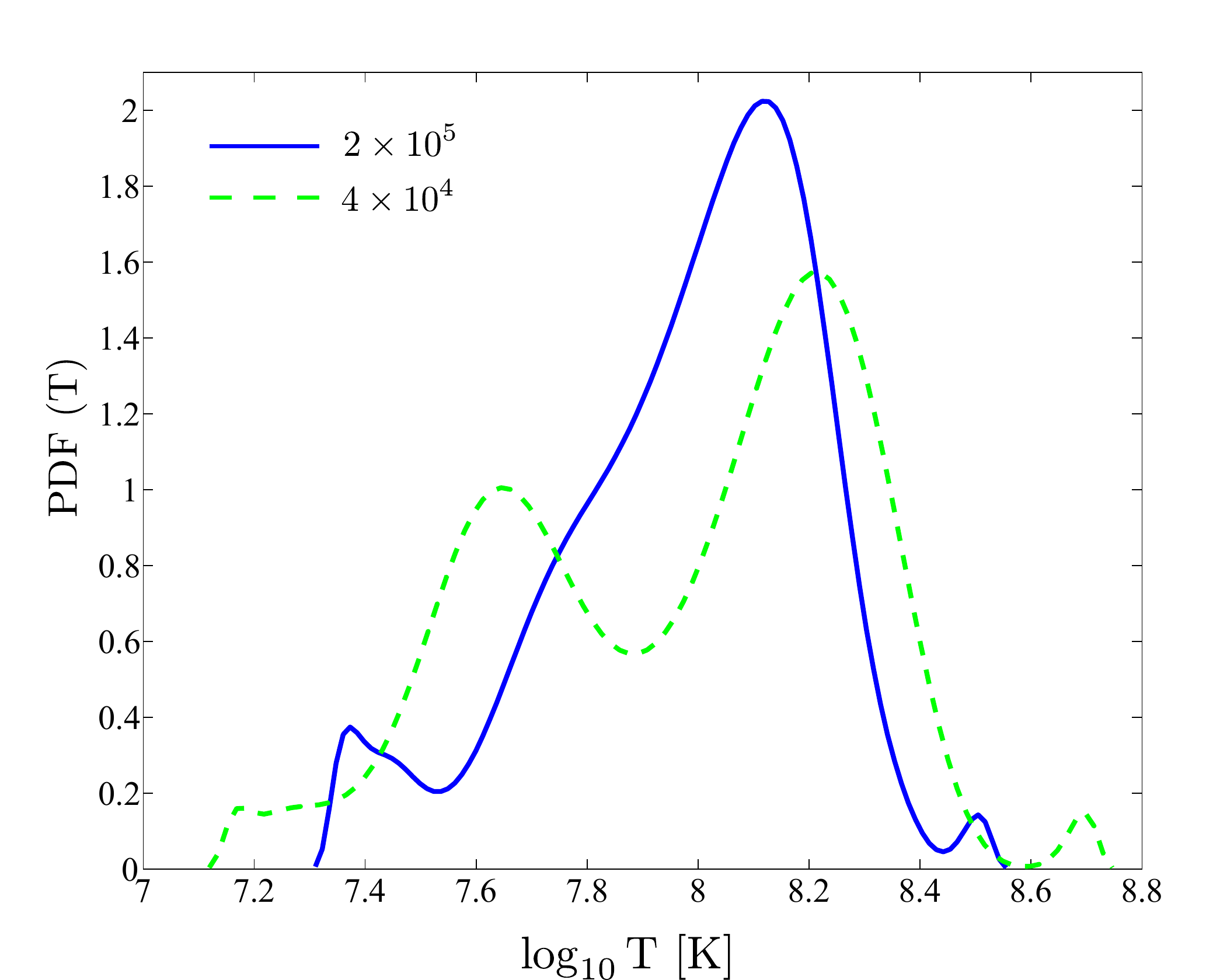}
  }
  \caption{Particle temperatures within a cross-section slice with a thickness
    equal with four smoothing lengths for a $0.3+0.6$ \Msun system using
    $4\times 10^4$ (left) and $2\times 10^5$ (center) SPH particles.
    The right panel shows the 
    probability distribution functions (PDF) of the temperature within the red
    circle (interaction radius of the particle with minimum $(\tau_{\rm
      nuc}/\tau_{\rm dyn})(T)$) shown in the left and central panels.}
  \label{fig:converg}
\end{figure*}

\subsection{Detonation criteria}
\label{sec:detcrit}

In its simplest form, a detonation is a shock wave that advances supersonically
into an unburnt, reactive medium. Behind the shock wave exothermic reactions 
take place that continuously drive the shock to wear down possible dissipative 
effects. Whether a detonation forms or not depends delicately on the exact local
conditions. The detonation conditions that have been applied in previous work
\citep{fink07,roepke07,townsley07, jordan08} ``all rather loosely decide whether
detonation conditions are reached based solely on the peak temperature reached
in the simulation above a certain density'' \citep{seitenzahl09}.
Apart from local density and temperature, the temperature
profile and the size of the region being heated determines whether a detonation is triggered or not. Depending on 
these factors, the critical radii have been found to vary between centimeters and hundreds 
of kilometers \citep{seitenzahl09}. In general, these values are  substantially below the
resolution lengths in the relevant regions  that we can afford in this study 
($\sim 10^7$ -- $10^8$ cm) with its more than 200 simulation runs.

Given the delicacy of the exact detonation conditions, we resort in the bulk of this
work to a simple yet robust timescale argument. We assume that dynamical burning 
sets in when the thermonuclear timescale, $\tau_{\rm nuc}$,  drops below the local 
dynamical timescale $\tau_{\rm  dyn}$.  When this happens matter heats up more 
rapidly than it can expand, cool and quench the burning. To compute $\tau_{\rm
  nuc}$ we use 
$\tau_{\rm  nuc}=c_p T/\epsilon_{\rm nuc}$ \citep[e.g.,][]{taam80,nomoto82}, 
where the nuclear energy generation rate, $\epsilon_{\rm nuc}$, is taken directly 
from the reaction network and $c_p$ is the specific heat at constant pressure, 
which we calculate from the Helmholtz equation of state. For the 
local dynamical timescale we use  $\tau_{\rm dyn}= H/c_s$, where $c_s$ is the sound 
speed and $H$ the local pressure scale height. We estimate the $H$ as $P/\rho g$, 
where $g$ is the local gravitational acceleration.

In our SPH formulation, the energy variable that is evolved forward in time is the
internal energy. The temperature at each particle position is found at every time step 
by a Newton-Raphson iteration. This temperature $T_a$, $a$ labelling the particle, 
is the best estimate for the local matter temperature, but small numerical fluctuations  
in the internal energy (say due to finite numerical resolution, finite SPH neighbour 
number, finite time stepping accuracy etc.) can lead to larger fluctuations in the 
SPH particle temperature estimate. This is especially so in regions where matter is 
degenerate and essentially independent of temperature. To some extent, this could be alleviated by designing a Fehlberg-type integration scheme that rejects 
and re-takes time steps if their temperature changes are too large. In addition to being computationally expensive, this 
approach may be ineffective in the case the fluctuations are not associated with the choice of time-step.
Therefore, one should be very cautious in interpreting temperatures of very few (substantially 
less than the typical neighbour number) hot particles, which may also be due to
small internal energy fluctuations.

Although we stress that  $T_a$ is the best local temperature estimate, 
we also calculate the SPH-smoothed temperature value
\begin{equation}
\langle T\rangle_a=\sum_b\frac{m_b}{\rho_b}T_bW(|\vec r_a-\vec r_b|,h_{a}),
\label{eq:tsmooth}
\end{equation}
which is a robust lower limit. Here,  $\rho_b$ and  
$m_b$ are the densities and masses of neighbour particles $b$, $W$ is the cubic spline 
kernel and $h_a$ the smoothing length of particle $a$.  In the limit of infinite resolution, both 
temperature estimates would coincide. Although it is
plausible that a detonation would 
be triggered when $(\tau_{\rm nuc}/\tau_{\rm dyn})(T) < 1$, we only consider a
detonation as inescapable consequence beyond reasonable doubt if the more
conservative ratio $(\tau_{\rm nuc}/\tau_{\rm dyn}) (\langle T\rangle)$ is also below
unity.

As a matter of illustration we show in Figure~\ref{fig:diffTrho} the evolution of temperature
and density of a $0.6+0.9$ \Msun system during the last stages of its merger. The left panel
shows the maximum particle temperature (green) together with the SPH-smoothed value
(red) and the arithmetic mean value within the interaction radius ($=2h$) of
the hottest particle (white).  
The shaded regions indicate the 1 and 2 $\sigma$ deviations. The corresponding quantities for
the densities are plotted in the right panel. 
To illustrate the geometry of the potential ignition region we show in
Figure~\ref{fig:converg} (left and 
central panel)
the particle temperatures within a slice around the orbital plane. For a fair comparison we chose
the slice thickness in each case to be 4 times the smoothing length of the particle with the
minimum ratio of $(\tau_{\rm nuc}/\tau_{\rm dyn})(T)$. The interaction 
region of this particle is indicated in both panels by the red circles. The
probability distribution functions 
(PDFs) of the particles in these interaction regions are shown in the right panel.
Despite the very low numerical resolution that can be afforded in this
parameter study,  the overall 
dynamics is captured accurately, see for example the excellent agreement of the numerical results
with the analytical estimates for the orbital evolution shown in
Figure~\ref{fig:norbs}. Nevertheless, at the current   
low resolution the shown interaction regions can contain particles from different morphological 
parts of the flow. In the shown example the most promising particle has neighbours in  the 
cold ($T<5 \times 10^7$ K) incoming accretion stream, in the hot interaction region 
($\sim 2 \times 10^8$ K) and in the moderately hot inner disk regions  ($\sim 10^8$ K). As a 
consequence the PDFs can be multi-peaked. As resolution increases and the neighbour
particles come from a smaller and smaller local volume, the PDFs will become  narrower,
single-peaked and $T$ and $\langle T \rangle$ will become increasingly
similar.

In the remainder of the paper we explore the $M_1-M_2$ plane to determine the regions 
in which the detonation conditions we have described are fulfilled.

\subsection{Helium mass transferring systems}
\label{sec:HeMT}

As discussed in Section~\ref{sec:ICs}, some of the donors we consider are composed of He in their outermost layers. For those systems with either He
cores or thick He shells, the mass transferred to the accretor will be pure He
until either the donor is disrupted, or until the He layer on the donor is
exhausted. As donors with a significant fraction of He may exist for $0.2
\leq M_{2} \leq 0.6$ \msun, the mass ratio of systems with He
donors can vary quite widely, and thus both dynamically stable and unstable
systems with He donors are expected to exist. Those dynamically stable systems
that are close enough to one another to initiate mass transfer will appear as
AM CVn systems, whereas the dynamically unstable
systems, which experience an exponential growth in the accretion rate, are only likely to be observed prior to contact. As WDs are supported
by degeneracy pressure, the mass-radius relationship of WDs only weakly
depends on the WDs' composition, and thus the ratio of He to CO in WDs is
largely unknown, aside from those WDs that are too massive to contain any
significant He.

\subsubsection{He detonations in stable systems}
Observed AM CVn systems are primarily expected to accrete through a disk
\citep{nelemans05}, although a few systems are likely undergoing direct impact
accretion \citep{marsh02,ramsay02}. While the temperature necessary for
dynamical He burning is 
somewhat higher than the virial temperature of a $\sim 0.6$ \Msun WD, slow
He burning can proceed in the accretor's acquired He shell over a timescale
shorter than the thermal timescale, raising the entropy of the accreted
material before the excess heat can be radiated away into space. This is in
fact the reason why no single WDs with a significant He layer are expected for
$M \gtrsim 0.6$ \msun, as even if the WD begins its life with a layer of
He, this layer will be converted into CO on a timescale that is significantly
shorter than the Hubble time. If the density at the base of the He layer is
large enough (i.e., if the He shell is massive enough), the temperature
eventually rises so much that the burning timescale becomes shorter than the
convective timescale in the envelope, leading to a run-away process that
produces a detonation \citep{bildsten07}. 

However, a detonation can only occur in these kinds of systems if the donor is
massive enough such that it can provide the minimum required mass. It turns
out that the minimum mass necessary is larger than $\sim 0.2$ \Msun for
accretors with mass $M_{1} < 0.6$ \Msun \citep{shen09}, systems that may or
may not be dynamically stable depending on the synchronization timescale. 

\begin{figure}
  \centerline{
    \includegraphics[height=2.5in]{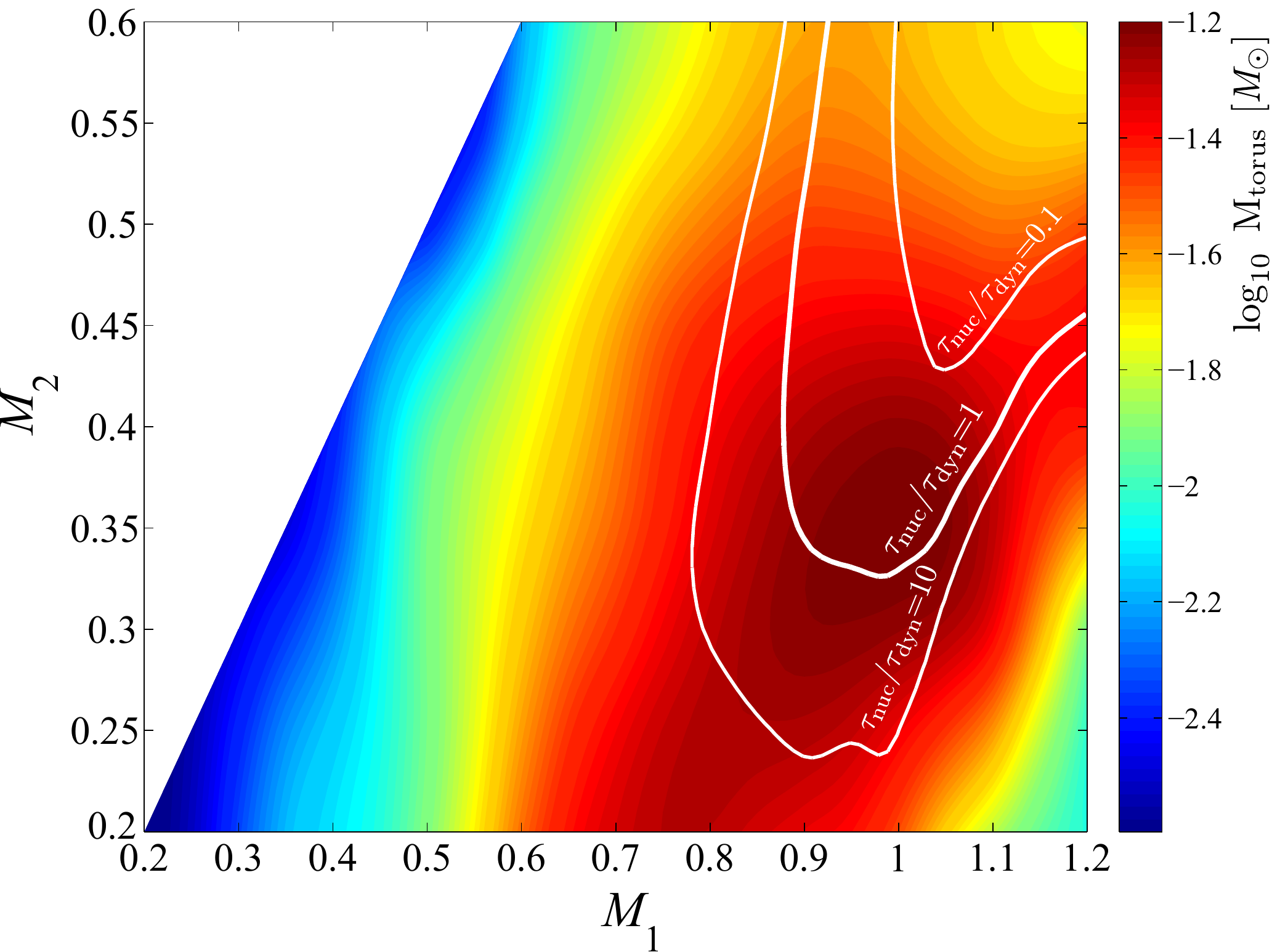}
  }
  \caption{Torus masses of the He donor systems. If ever $\tau_{\rm nuc} \le
    \tau_{\rm dyn}$, we measure the masses when the timescales for the first
    time become equal, otherwise we take the mass when the ratio $\tau_{\rm
      nuc}/\tau_{\rm dyn}$ is minimal.
    The white contours show $\tau_{\rm
  nuc}/\tau_{\rm dyn}$ when the ratio is at a minimum, with the thick contour
showing where the ratio is unity. 
A comparison with Figure~\ref{fig:detonations} shows that a large fraction of
WDs with He donors undergo accretion stream triggered He-detonations prior to
contact. 
The torus masses at detonation can be as large as
$\sim 0.1$ \msun, but also significantly smaller for larger $M_{\rm
  tot}$. Systems with a massive He donor and small disk masses at detonation
may be capable of producing multiple sub-luminous events prior to final
coalescence. }
  \label{fig:streamdet}
\end{figure}

\subsubsection{Stream-induced detonations prior to contact}
As we described in \cite{guillochon10}, detonations of the He torus can
be triggered by the accretion stream several orbits prior to the merger
event. Using our significantly larger set of accreting systems, we have
provided an updated version of the figure originally presented in \cite{dan11} showing the systems that are
expected to have surface detonations triggered by the accretion stream, see
Figure~\ref{fig:streamdet}. A thermonuclear runaway will occur
prior to the merging event if $\tau_{\rm nuc}/\tau_{\rm dyn}
\sim 1$, a condition that depends on the thermodynamic state of the He
torus acquired through the accretion process. The conditions for burning are
more easily met for systems in which the temperature and density of the
material acquired pre-merger is large to begin with, leading to conditions
favorable for uncontrolled burning once the material is compressed by the
dynamical interaction, whether by the stream or by the donor itself. 

The pre-interaction conditions in the He torus depend on three primary
factors: the mass of the accretor, the component of the stream velocity normal
to the accretor's surface at impact, and the evolution of $\dot{M}$ with
time. For increasing accretor mass, the larger gravitational pressure at the
accretor's surface leads to larger hydrostatic pressures, and thus greater
equilibrium densities and temperatures. Conversely, an increasing accretor
mass shrinks the accretor's radius relative to the circularization radius,
resulting in more ``disk-like'' accretion and thus a 
less-concentrated torus that is more supported by rotation than fluid
pressure. A shrinking accretor also reduces the normal component of the stream
velocity relative to the accretor's surface, and as a result the degree of
compression achieved as a result of the stream's impact onto the accretor's
surface is more limited. These effects are manifest as the larger $\tau_{\rm
  nuc}/\tau_{\rm dyn}$ ratios toward the highest accretor masses in Figure~\ref{fig:streamdet}. 

A necessary condition for successful stream detonations is that the stream's
ram pressure is considerably larger than the hydrostatic pressure of the
torus, which is only possible if the mass of the stream is comparable in mass
to the torus that has already accumulated. This implies that the {\it second}
derivative of the amount of mass transferred, $\ddot{M}$, must be comparable
to $\dot{M}$ itself. Therefore, stream detonations are only possible when an
exponentially growing $\dot{M}$ can be sustained for several orbits, a
condition that is typical of systems with abrupt orbital evolution. Because of the stochastic nature of the triggering event, a detonation can be realized at any point in time where the He torus mass is
larger than the minimum shell mass required for a successful detonation
\citep{bildsten07,shen09,woosley11}, which we find can precede the final merger event by
many orbits. 

\subsubsection{Contact detonations}

\begin{figure*}
  \includegraphics[height=6.5in]{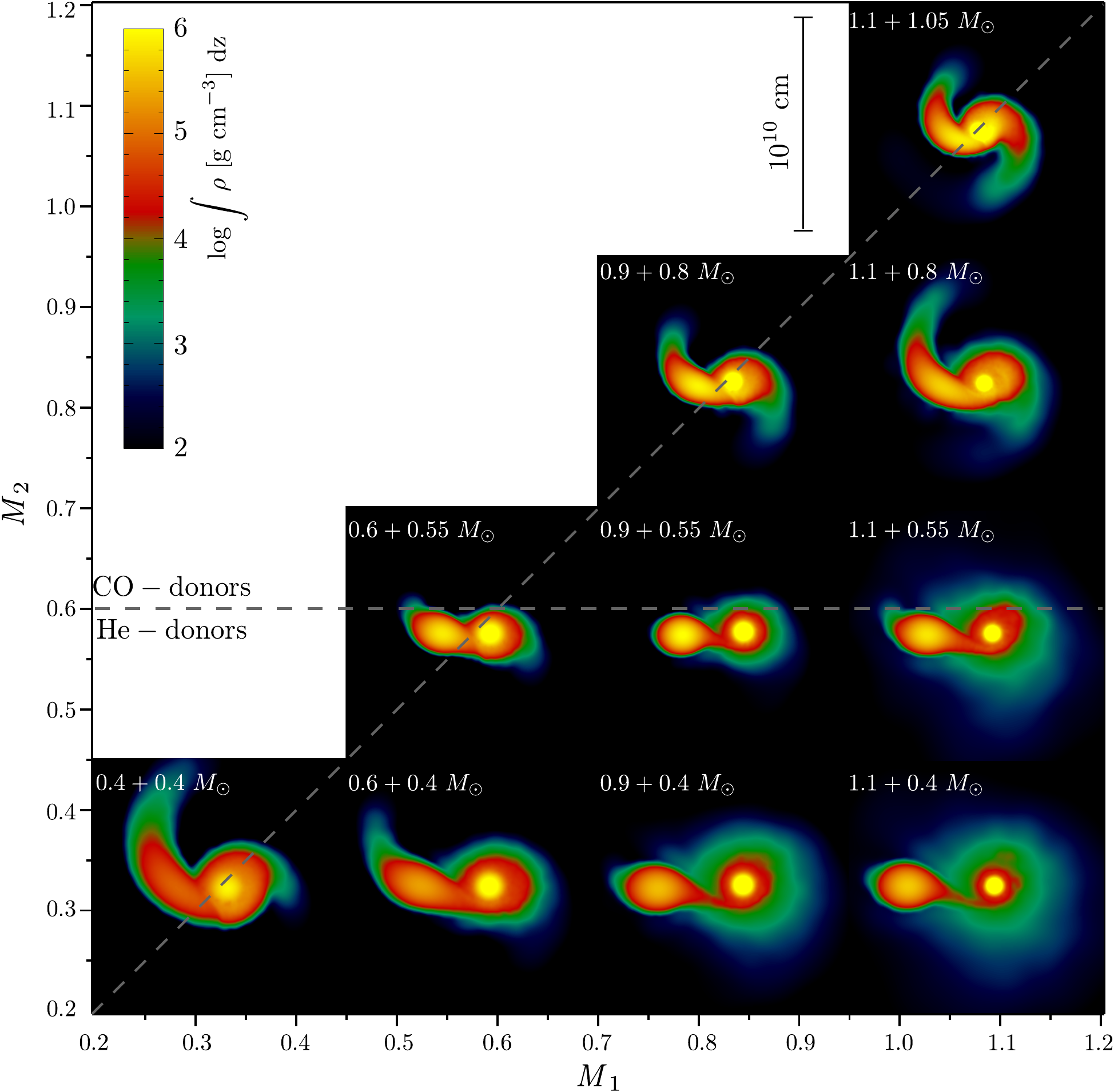}
  \caption{Column density snapshots corresponding to the moment when the
    system first crosses $(\tau_{\rm nuc}/\tau_{\rm dyn})(T)<1$, or, if that
    does not occur, when $(\tau_{\rm nuc}/\tau_{\rm dyn})(T)$ is minimum. The ten
    snapshots shown here are representative of all DD chemical composition and mass
    combinations shown in Figure~\ref{fig:comp}. In this
    parameter space we fully explore the possible 
    combinations of WD donors and accretors and investigate their
    orbital stability and whether detonations prior or at the merger moment
    are possible.
    Systems with a mass ratio close to one (diagonal dashed-line) show a
    quick disruption, within about 20 orbits, while those away from this
    line show a slow 
    depletion of the donor (ie. low $\dot M$) over dozens of orbits of mass
    transfer and the formation of an extended atmosphere surrounding the
    accretor.
    We find that a large fraction of the systems with 
    He donors (below the horizontal dashed-line) are expected to
    explode prior to or at the point of merger. In contrast, the CO
    accreting systems (above the horizontal dashed-line) are unlikely to
    explode at or prior to the merger.}  
  \label{fig:morphology}
\end{figure*}

If detonations from the stream interaction are avoided, the system has another
chance to initiate a He detonation during the final coalescence of the two
systems. Except for WDs that are almost equal in mass, the dynamical
interaction during the final coalescence involves the donor being ripped apart
by tides, while the accretor remains relatively unaffected. Approximately half
of the donor's remaining mass falls on top of the previously accreted He, but
at a speed that's only a fraction of the escape speed from the accretor as the
shredded donor is partly held aloft by its angular momentum (Figure~\ref{fig:morphology}). Thus the final
coalescence can be viewed as a low-speed collision \citep{rosswog09b}. In
simulations of  WD-WD head-on collisions, collisions between pairs
of WDs with mass as low as 0.4 \Msun He WDs were seen to produce
detonations. The mutual escape speed of two 0.4 \Msun at the point of
collision is $\sqrt{2 G (M_{1} + M_{2}) / (R_{1} + R_{2})} = 3 \times 10^{8}$
cm s$^{-1}$, which is approximately one-third of the escape speed from a $0.8$
\Msun accretor. 

\begin{figure*}
  \centerline{  
    \includegraphics[height=2.3in]{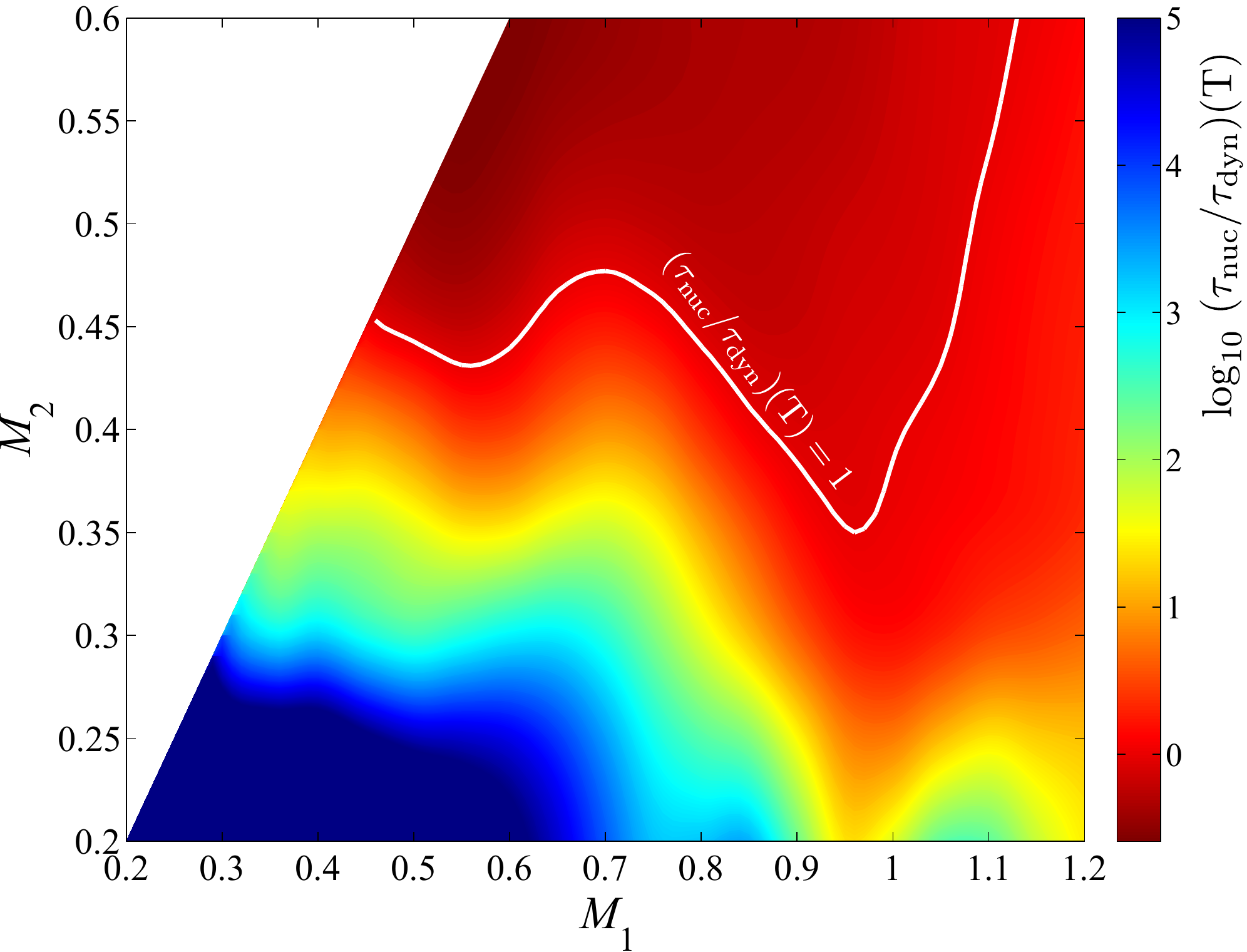}\hspace{0.5cm}
    \includegraphics[height=2.3in]{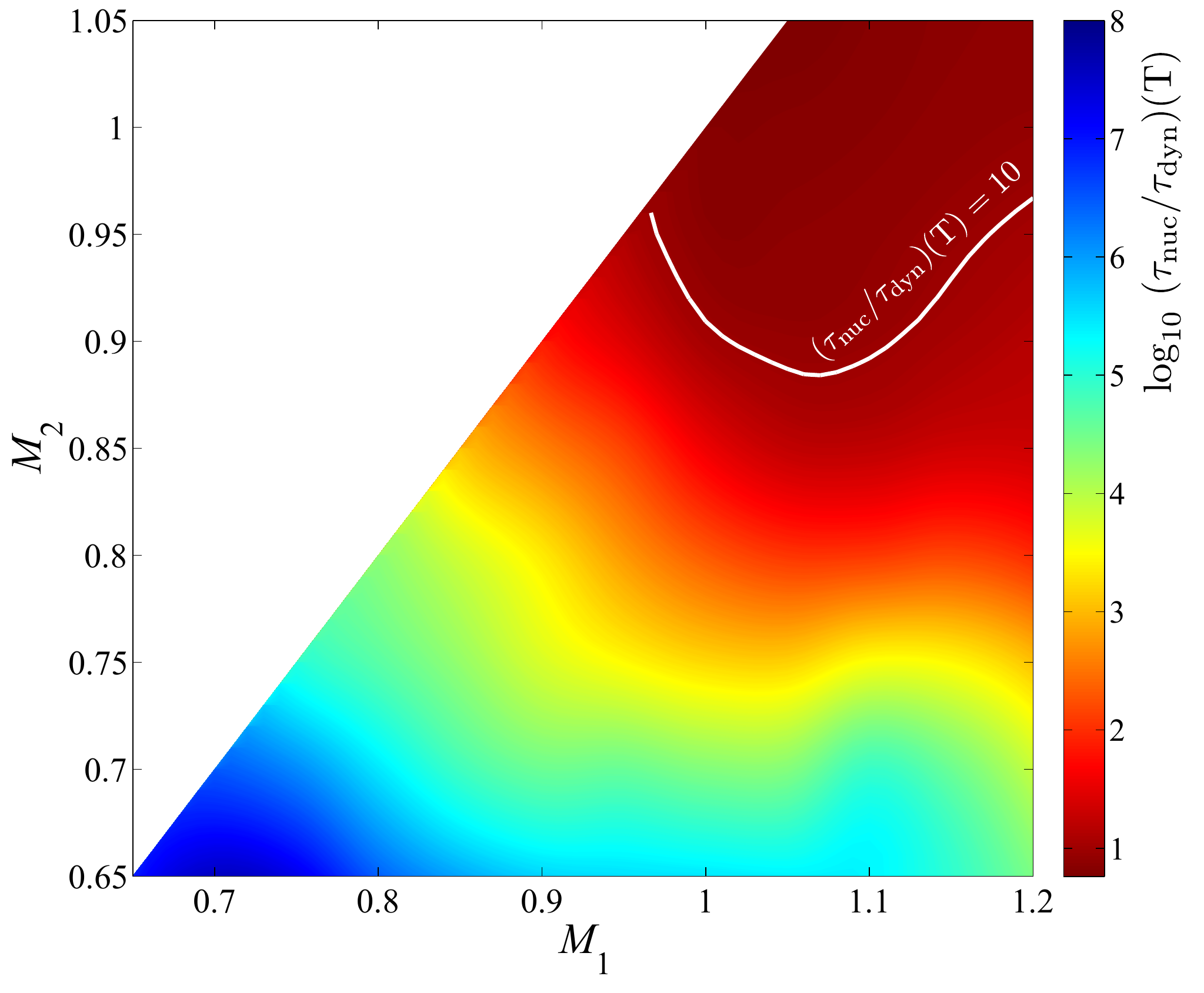}}
  \centerline{  
    \includegraphics[height=2.3in]{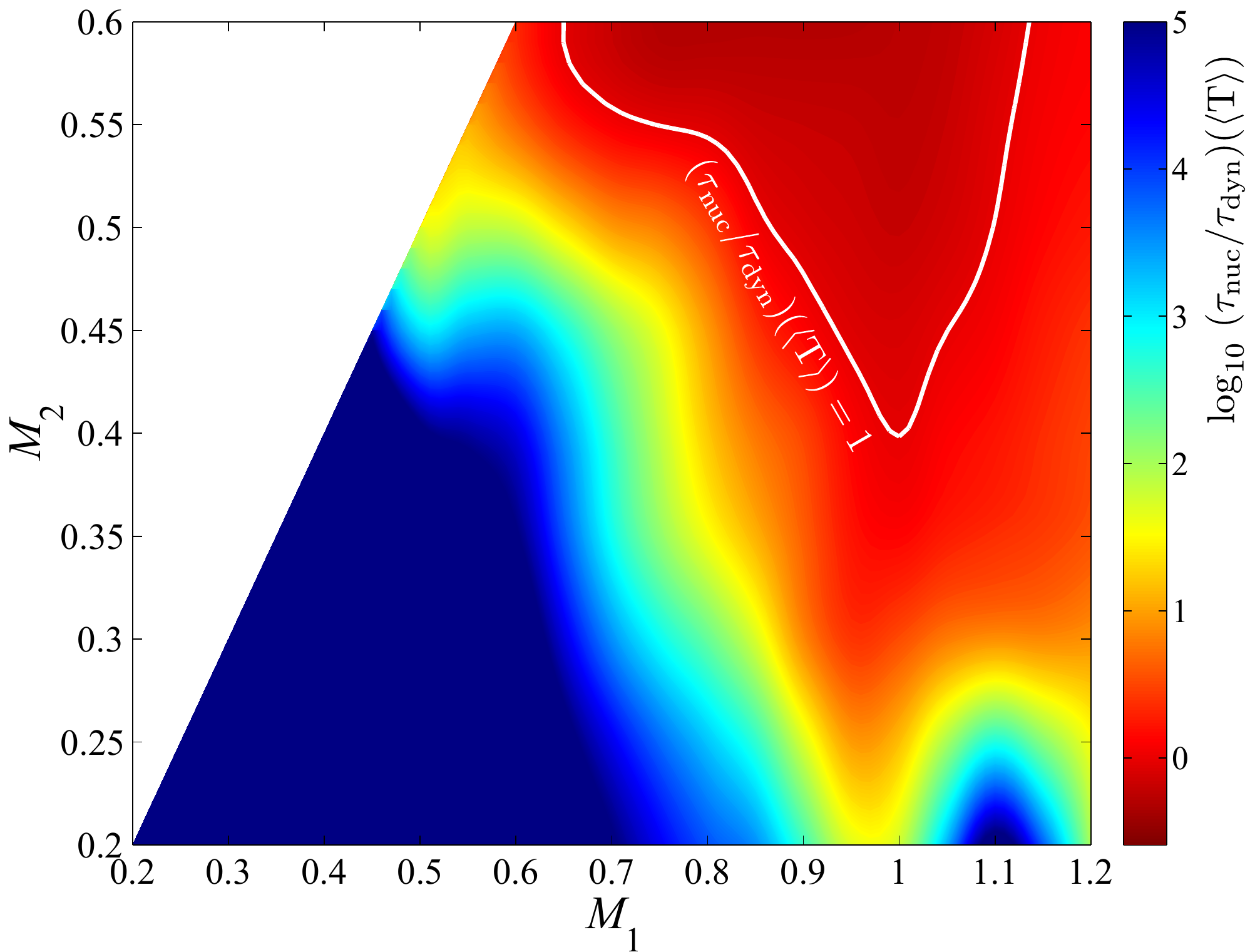}\hspace{0.5cm}
    \includegraphics[height=2.3in]{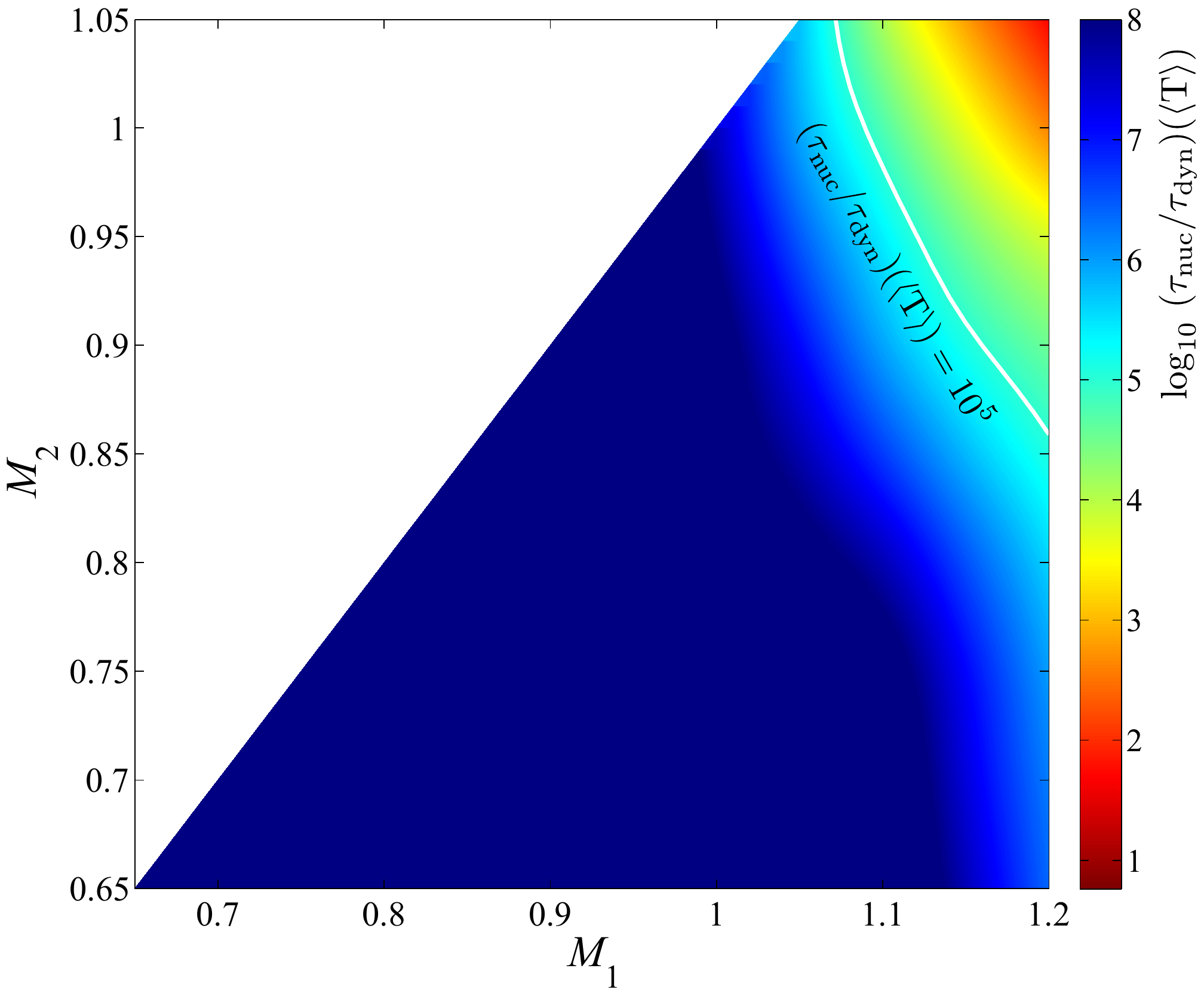}
  }
  \caption{Logarithmic ratio of the thermonuclear timescale $\tau_{\rm
      nuc}$
    to the local dynamical timescale $\tau_{\rm dyn}$ at the moment of merger over a grid spanning
    the entire parameter space of He (left) and CO donors (right)
    masses. The ratio was computed using both the
    individual particle temperature $T$ (upper panels) and the SPH-smoothed
    temperature $\langle T\rangle$ (lower panels, see eq. \ref{eq:tsmooth}). Thermonuclear runaway is
    expected for configurations above the contours with $\tau_{\rm
      nuc}/\tau_{\rm dyn}=1$.  
  }
  \label{fig:detonations}
\end{figure*}

At the point of disruption, half of the donor star falls onto the accretor as
the donor's Roche lobe rapidly shrinks. Once this has occurred, the
donor star can be viewed as material whose orbital dynamics are primarily
determined by the surviving accretor. This means that the common center of
motion, the barycenter, shifts from being located at a point lying between the
two WDs to lying within the center of the accretor. From the plunging donor
material's point of view, which originally moved on a circular orbit, the
point around which it orbits moves, placing the material on a highly eccentric
orbit with a large $\dot{r}$. This results in a collision that is more
violent, with a plunging speed that can be almost as large as the accretor's
escape speed. In systems with a large $q$, the barycenter is usually already
close to the accretor's center of mass, and thus there is no rapid acceleration within the corotating frame, and thus the final coalescence is much more gentle. This
behavior is clearly visible in the right panel of Figure~\ref{fig:detcritCO},
which shows that the temperature at minimum $(\tau_{\rm nuc}/\tau_{\rm
  dyn})(T)$ tends to increase with 
increasing $q$ for a fixed total mass. 

The optimal conditions for ignition are reached when the
temperature is at its peak, close to the time of the donor's disruption, provided that the accretor is below 1.1 \Msun and the donor is above 0.4 \msun, see the upper-left panel of Figure~\ref{fig:detonations}. For all systems with $M_{\rm tot} < 0.9 M_{\odot}$, individual particle temperatures $T$ of $10^8$ K and greater are reached in the torus formed via the mass transfer, but the density is too low to lead to a dynamical burning event. If we consider the more conservative smoothed temperature criteria, only systems with $M_{\rm tot} > 1.3 M_{\odot}$ are capable of detonating at contact. Under both criteria, systems with $q \sim 1$ are more prone to produce detonations. Systems that do not satisfy the criteria for detonating either by the stream mechanism or at the point of merger may be progenitors of He-novae instead \citep[e.g.,][]{yoon04}.

The ratio $\tau_{\rm nuc}/\tau_{\rm dyn}$ is very large in the right-lower
corner of Figure~\ref{fig:detonations}, left panels, close to the region of
disk accretion. As described in Section~\ref{sec:orbstab}, we expect that the systems should be dynamically stable, as they do not show any sign of a merger
after $70-90$ orbits of mass transfer.

\subsection{Carbon-Oxygen mass transfer}
\label{sec:COMT}

\begin{figure*}
  \centerline{
    \includegraphics[height=2.5in]{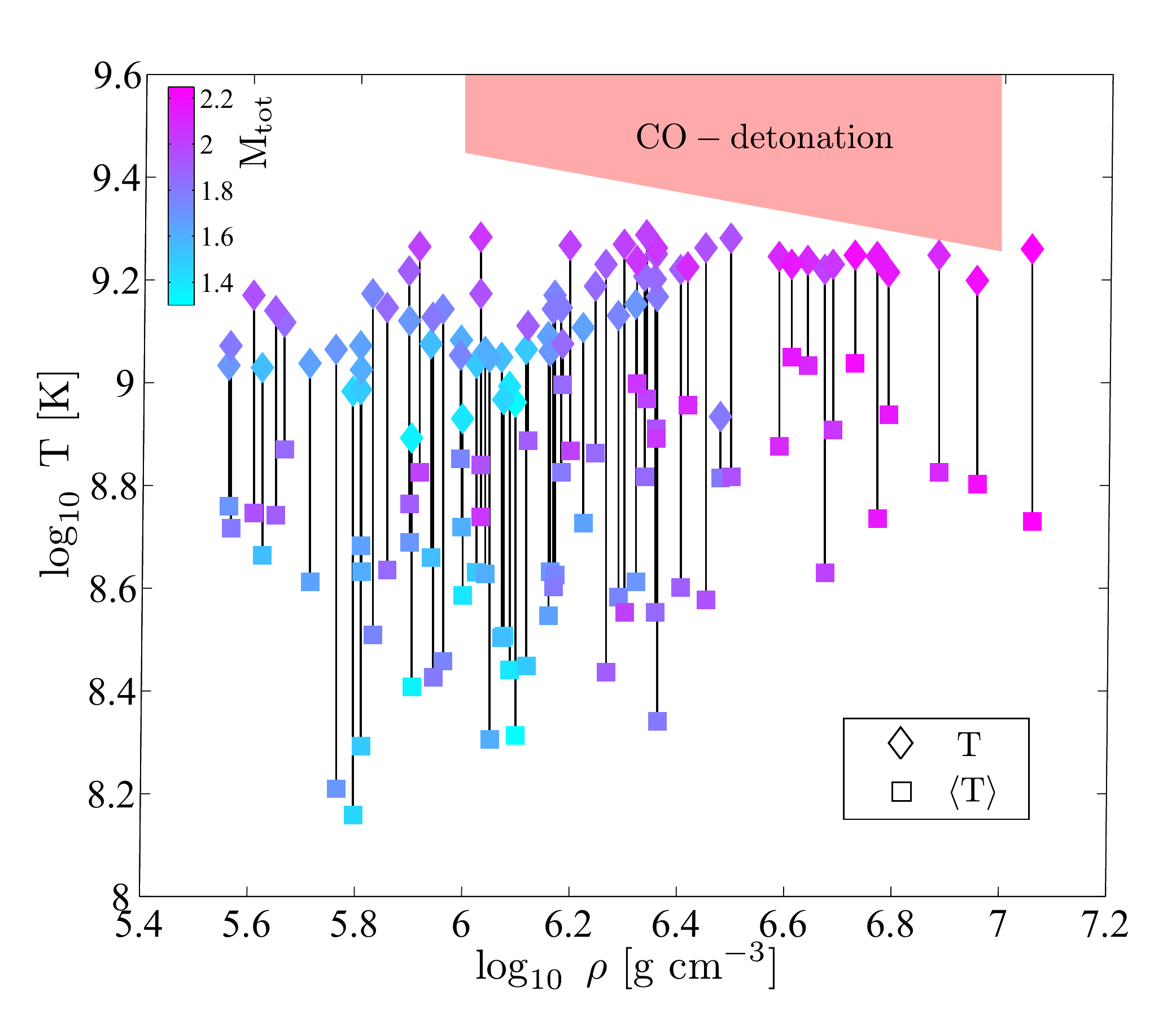}\hspace{0.8cm}
    \includegraphics[height=2.5in]{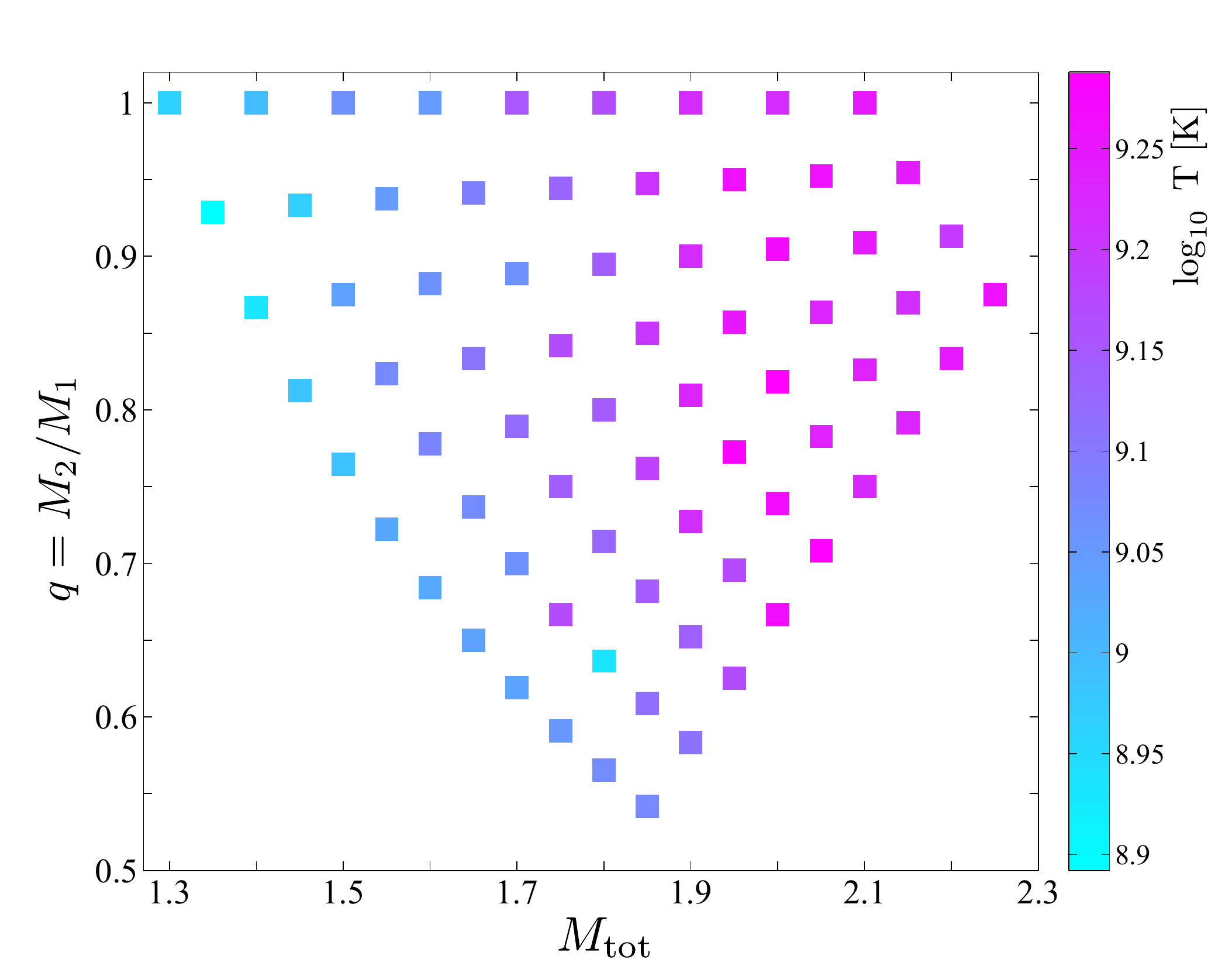}
  }
  \caption{Left panel:
   The filled diamonds show $(\rho,T)$ at minimum $(\tau_{\rm nuc}/\tau_{\rm
     dyn})(T)$ and are connected by black lines to the $(\rho,\langle
   T\rangle)$, represented by filled squares. Color coded is
   the total mass of the system in solar units. $\langle T\rangle$ is
   computed at the location of particle with temperature $T$. None of the CO mass-transferring systems reaches the CO detonation 
   criteria of \citet{seitenzahl09} (shaded region) or the $\tau_{\rm nuc}/\tau_{\rm dyn}<1$ condition (see the right panels of Figure
   \ref{fig:detonations}). Right panel:
   Temperature at minimum  
   $(\tau_{\rm nuc}/\tau_{\rm dyn})(T)$ in $q - M_{\rm tot}$ plane.
   The temperature shows a clear trend towards higher values with increasing mass
   ratio and/or fixed total mass. }
  \label{fig:detcritCO}
\end{figure*}

While we find that He accreting WDs yield detonations, systems accreting CO are
unlikely to explode at or prior to merger, see the right panels of Figure~\ref{fig:detonations}. 
Figure~\ref{fig:detcritCO} shows the region of $\rho-T$ plane where a CO
detonation can be successfully ignited \citep{seitenzahl09}, together with the
temperature, both $T$ and $\langle T\rangle$, and density of all CO
mass-transferring systems at the moment when 
$(\tau_{\rm nuc}/\tau_{\rm dyn})(T)$ is minimal.
Overall, there is a clear tendency to increase the temperature at minimum $(\tau_{\rm
  nuc}/\tau_{\rm dyn})(T)$ with the mass
ratio for a fixed total mass, see the right-panel of Figure~\ref{fig:detcritCO}. 

Based on the detonation criteria of
\cite{seitenzahl09}, \cite{pakmor11} have found detonations in their 
merger study of (nearly) equal mass WDs.
Because \cite{pakmor11} used approximate ICs that result
in higher temperatures and densities during the evolution of mass transfer
\citep{dan11}, we run a direct comparison with their $0.81 + 0.9$ \Msun
system. We compared the minimum obtained value of $\tau_{\rm nuc}/\tau_{\rm dyn}$ with
accurate \citep{dan11} and approximate \citep[same as in][]{pakmor11} ICs for
non-rotating stars and found a difference of only $5\%$, with $\tau_{\rm
  nuc}/\tau_{\rm dyn}\approx 7$.
The difference is much larger however, for systems with lower mass ratios. For
the corotating system with $0.6 + 0.9$ \Msun components, $\tau_{\rm 
  nuc}/\tau_{\rm dyn}$ for the approximate ICs is four orders of magnitude
greater than the value obtained for using the accurate ICs.
The two ICs also lead to important differences in the merger remnant profiles
at the moment when $\tau_{\rm nuc}/\tau_{\rm dyn}$ is minimum: the temperature
is overestimated when using the approximate ICs and the density is larger in
the hot envelope surrounding the accretor, where the dynamical burning takes
place (Figure~\ref{fig:compICs081}).

\section{Discussion}
\label{sec:discussion}

Although small patches of the white dwarf merger parameter space  have been
explored previously in a moderate number of simulations 
\citep{benz90,rasio95,segretain97,guerrero04,dsouza06,yoon07,motl07,aguilar09,pakmor10,fryer10,pakmor11,dan11,raskin11},  
the basic question: {\it Which white dwarf systems produce explosions prior to or directly 
at merger?} has so far remained unanswered. This serious gap in our understanding
has in particular hampered our ability to confidently judge the potential of
DD mergers as type Ia progenitor systems. In this paper we close this
gap by methodically covering the  
whole parameter space of white dwarf mergers using more  than 200 simulation runs.
We have systematically scanned the white dwarf mass range from 0.2 to 1.2 \Msun
and accounted for their different initial chemical compositions.

\begin{figure*}
  \centerline{
  \includegraphics[height=2.8in]{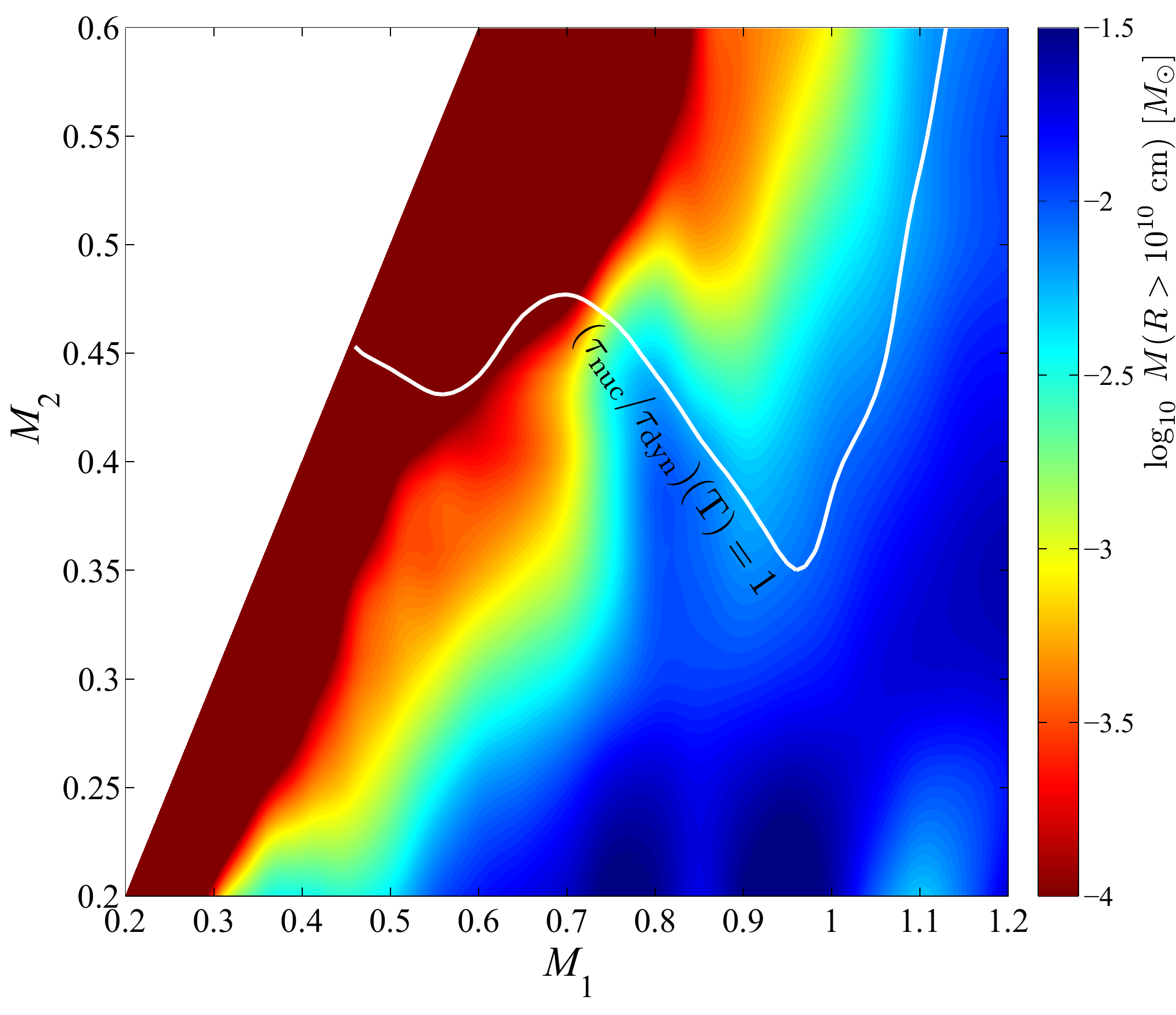}\hspace{0.5cm}
  \includegraphics[height=2.8in]{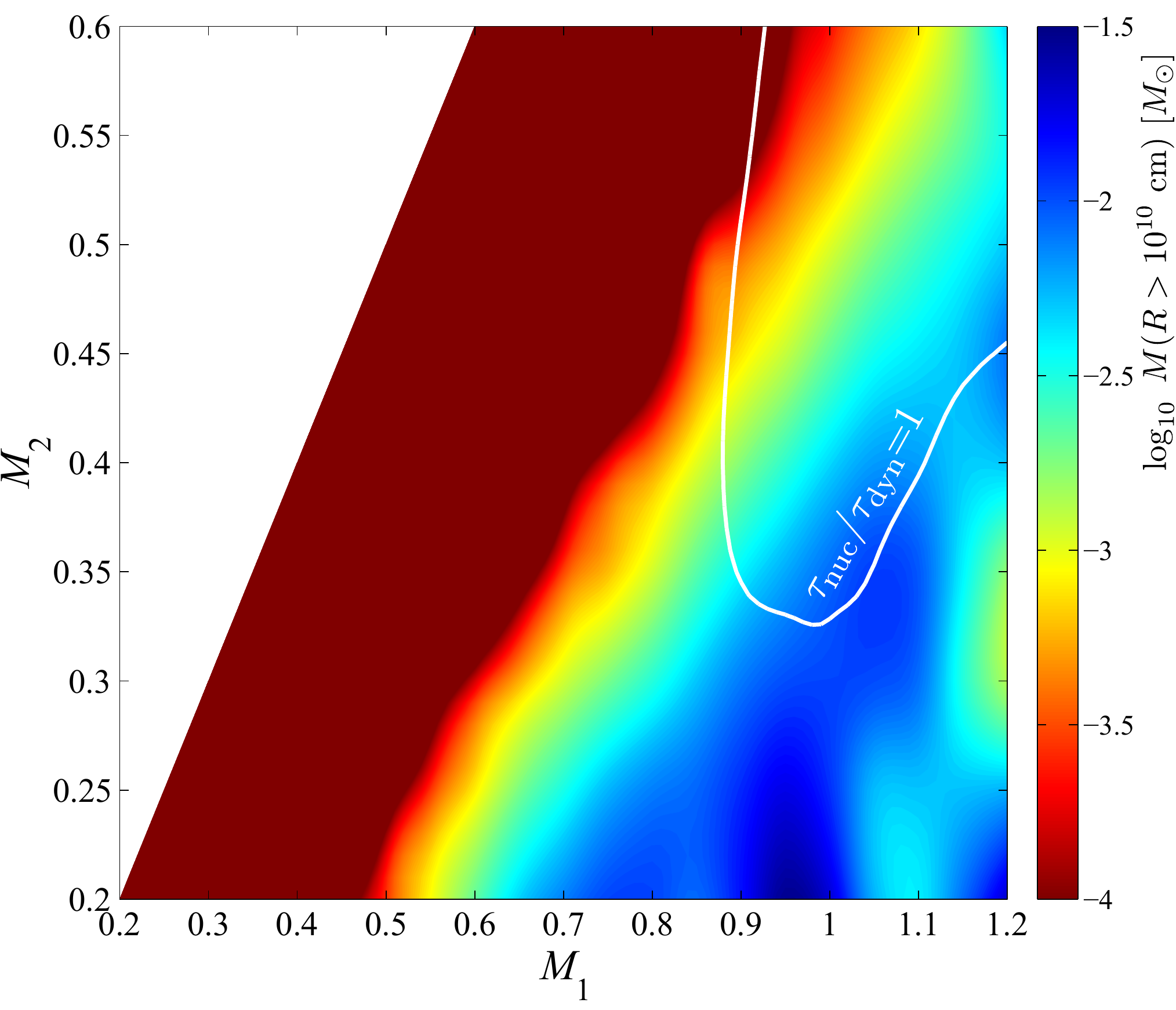}
  }
  \caption{Mass (in solar units) above a radius of $10^{10}$ cm with respect to the
    accretor's center of mass for contact- (left) and stream-induced (right) He
    detonations. As in Figures~\ref{fig:streamdet} and \ref{fig:detonations}, we calculate the masses when the timescales for the first
    time become equal, or, if that does not occur, we take the mass when the
    ratio $\tau_{\rm nuc}/\tau_{\rm dyn}$ is minimal.
    The white contours show where the ratio $\tau_{\rm
  nuc}/\tau_{\rm dyn}$ is unity.}
  \label{fig:mabove_1e10}
\end{figure*}

While such a broad study is necessarily limited in the numerical resolution that 
can be afforded in each simulation, we have found excellent qualitative agreement 
with previous numerical studies \citep{dan11} and semi-analytical work 
\citep{marsh04} on the orbital stability of white dwarf binary systems. 
One of the main conclusions of our previous study \citep{dan11}
was that the accuracy of the numerical initial conditions has a major impact on all 
subsequent evolutionary stages and can be more important than numerical resolution.
All our simulations were started from
carefully constructed, initially tidally-locked binary configurations.
The dynamically unstable systems in nature should --at the initial
separations where numerically resolvable mass transfer sets in-- have an 
accretor star that is desynchronized with the orbit \citep{new97}. In practice, these initial 
conditions require a cumbersome approach as desynchronicity is a strong de-stabilizing 
effect, and therefore constructing a relaxation scheme that does not lead to a premature 
merger is difficult. However, in all of our simulations the accretor star becomes desychronized 
with the orbital period after about 5 to 10 orbital revolutions. We find systems that are  
near the dividing line between disk and direct impact accretion can 
still persist for many dozens of orbits prior to merging, despite the fact that the accretor 
is already desynchronous. We also find that as the binary approaches merger,
the orbital period changes so quickly that corotation cannot be
maintained. In other words, 
while our initial conditions do not represent the physical state at the onset of resolvable 
mass transfer, the systems quickly evolve to the state that is likely realized in nature.

As expected, systems of a given mass ratio lead to higher peak temperatures as the total system
mass increases. We additionally find that there is a clear trend to produce higher peak temperatures 
as the mass ratio approaches unity (Figure~\ref{fig:detcritCO}).
It comes as an interesting and maybe somewhat surprising result that a large fraction
of the He-accreting systems are expected to explode prior to merger or at surface contact:
we find that all systems with accretor masses below 1.1 \Msun undergo such explosions, 
provided that the He-donating white dwarf exceeds $\sim 0.4$ \msun (Figure~\ref{fig:detonations}).  This opens up a large fraction of  unexplored parameter
space for thermonuclear explosions.  
A substantial fraction of these systems could even explode earlier via 
stream-induced detonations, as we have demonstrated in our earlier work \citep{guillochon10}. 
DD binaries made entirely of carbon and oxygen, in contrast,
are unlikely to explode prior to or during the merger. This conclusion may need to be 
altered for cases where the CO white dwarfs possess more than a critical
amount of He in their outer layers \citep{shen09,raskin11}.  
Systems that do not undergo an early thermonuclear event may still be prone to an
explosion long after the merger \citep{yoon04}.

Based on a previous calculation \citep{fryer10}, it has been argued that the environment
surrounding exploding DD systems is polluted with tens of percent of a solar mass and 
would have produced a detectable optical excess during the early evolution of 2011fe 
\citep{nugent11}. However, in none of our exploding systems do we find more than 
$1\%$ of a solar mass exterior to $10^{10}$ cm, and in fact many of our systems have 
no measurable mass beyond that distance (Figure~\ref{fig:mabove_1e10}). Because we are starting with initially 
synchronized WD binaries, the total amount of angular momentum in our simulations 
is a robust upper limit. Therefore we can conclude that the ambient environment will 
contain even less mass at these distances than what our current simulations predict.

We have shown that DD mergers provide a broad channel for producing
thermonuclear explosions. While our simulations indicate that systems
undergo detonations within the He layer, it remains an open question
whether this primary detonation can actually trigger the explosion of the CO
core. In the event that the CO core fails to detonate, these events may still
resemble sub-luminous type Ia \citep{shen10}. Future research has to investigate
which of these thermonuclear explosions are consistent with current type Ia SN observations.

\section*{Acknowledgments}

We thank Philipp Podsiadlowski and Ken Shen for insightful comments and stimulating
discussions. We acknowledge support 
from DFG grant RO 3399 (M.D. and S.R.), the David and Lucile Packard
Foundation (J.G., S.R., E.R.-R.), NSF grant: AST-0847563 (E.R.-R.), and the NASA
Earth and Space Science Fellowship (J.G.).

\bibliographystyle{mn2e}
\bibliography{detonating_DWDs}

\begin{thebibliography}{90}
\expandafter\ifx\csname natexlab\endcsname\relax\def\natexlab#1{#1}\fi

\bibitem[{{Balsara}(1995)}]{balsara95}
{Balsara} D.~S., 1995, Journal of Computational Physics, 121, 357

\bibitem[{{Benz} {et~al}\mbox{.}(1990){Benz}, {Cameron}, {Press}, \&
  {Bowers}}]{benz90}
{Benz} W., {Cameron} A.~G.~W., {Press} W.~H., {Bowers} R.~L., 1990, ApJ, 348,
  647

\bibitem[{{Bildsten} {et~al}\mbox{.}(2007){Bildsten}, {Shen}, {Weinberg}, \&
  {Nelemans}}]{bildsten07}
{Bildsten} L., {Shen} K.~J., {Weinberg} N.~N., {Nelemans} G., 2007, ApJL, 662,
  L95

\bibitem[{{Bloom} {et~al}\mbox{.}(2011){Bloom}, {Kasen}, {Shen}, {Nugent},
  {Butler}, {Graham}, {Howell}, {Kolb}, {Holmes}, {Haswell}, {Burwitz},
  {Rodriguez}, \& {Sullivan}}]{bloom11}
{Bloom} J.~S. {et~al.}, 2011, ArXiv e-prints

\bibitem[{{Brown} {et~al}\mbox{.}(2011){Brown}, {Dawson}, {de Pasquale},
  {Gronwall}, {Holland}, {Immler}, {Kuin}, {Mazzali}, {Milne}, {Oates}, \&
  {Siegel}}]{brown11}
{Brown} P.~J. {et~al.}, 2011, ArXiv e-prints

\bibitem[{{Chomiuk} {et~al}\mbox{.}(2012){Chomiuk}, {Soderberg}, {Moe},
  {Chevalier}, {Rupen}, {Badenes}, {Margutti}, {Fransson}, {Fong}, \&
  {Dittmann}}]{chomiuk12}
{Chomiuk} L. {et~al.}, 2012, ArXiv e-prints

\bibitem[{{Clayton} {et~al}\mbox{.}(2007){Clayton}, {Geballe}, {Herwig},
  {Fryer}, \& {Asplund}}]{clayton07}
{Clayton} G.~C., {Geballe} T.~R., {Herwig} F., {Fryer} C., {Asplund} M., 2007,
  ApJ, 662, 1220

\bibitem[{{Dan} {et~al}\mbox{.}(2011){Dan}, {Rosswog}, {Guillochon}, \&
  {Ramirez-Ruiz}}]{dan11}
{Dan} M., {Rosswog} S., {Guillochon} J., {Ramirez-Ruiz} E., 2011, ApJ, 737, 89

\bibitem[{{D'Souza} {et~al}\mbox{.}(2006){D'Souza}, {Motl}, {Tohline}, \&
  {Frank}}]{dsouza06}
{D'Souza} M.~C.~R., {Motl} P.~M., {Tohline} J.~E., {Frank} J., 2006, ApJ, 643,
  381

\bibitem[{{Fink} {et~al}\mbox{.}(2007){Fink}, {Hillebrandt}, \&
  {R{\"o}pke}}]{fink07}
{Fink} M., {Hillebrandt} W., {R{\"o}pke} F.~K., 2007, A\&A, 476, 1133

\bibitem[{{Fryer} {et~al}\mbox{.}(2010){Fryer}, {Ruiter}, {Belczynski},
  {Brown}, {Bufano}, {Diehl}, {Fontes}, {Frey}, {Holland}, {Hungerford},
  {Immler}, {Mazzali}, {Meakin}, {Milne}, {Raskin}, \& {Timmes}}]{fryer10}
{Fryer} C.~L. {et~al.}, 2010, ApJ, 725, 296

\bibitem[{{Gil-Pons} \& {Garc{\'{\i}}a-Berro}(2001)}]{gilpons01}
{Gil-Pons} P., {Garc{\'{\i}}a-Berro} E., 2001, A\&A, 375, 87

\bibitem[{{Gokhale} {et~al}\mbox{.}(2007){Gokhale}, {Peng}, \&
  {Frank}}]{gokhale07}
{Gokhale} V., {Peng} X.~M., {Frank} J., 2007, ApJ, 655, 1010

\bibitem[{{Guerrero} {et~al}\mbox{.}(2004){Guerrero}, {Garc{\'{\i}}a-Berro}, \&
  {Isern}}]{guerrero04}
{Guerrero} J., {Garc{\'{\i}}a-Berro} E., {Isern} J., 2004, A\&A, 413, 257

\bibitem[{{Guillochon} {et~al}\mbox{.}(2010){Guillochon}, {Dan},
  {Ramirez-Ruiz}, \& {Rosswog}}]{guillochon10}
{Guillochon} J., {Dan} M., {Ramirez-Ruiz} E., {Rosswog} S., 2010, ApJL, 709,
  L64

\bibitem[{{Han} {et~al}\mbox{.}(2000){Han}, {Tout}, \& {Eggleton}}]{han00}
{Han} Z., {Tout} C.~A., {Eggleton} P.~P., 2000, MNRAS, 319, 215

\bibitem[{{Hicken} {et~al}\mbox{.}(2009{\natexlab{a}}){Hicken}, {Challis},
  {Jha}, {Kirshner}, {Matheson}, {Modjaz}, {Rest}, {Wood-Vasey}, {Bakos},
  {Barton}, {Berlind}, {Bragg}, {Brice{\~n}o}, {Brown}, {Caldwell}, {Calkins},
  {Cho}, {Ciupik}, {Contreras}, {Dendy}, {Dosaj}, {Durham}, {Eriksen},
  {Esquerdo}, {Everett}, {Falco}, {Fernandez}, {Gaba}, {Garnavich}, {Graves},
  {Green}, {Groner}, {Hergenrother}, {Holman}, {Hradecky}, {Huchra},
  {Hutchison}, {Jerius}, {Jordan}, {Kilgard}, {Krauss}, {Luhman}, {Macri},
  {Marrone}, {McDowell}, {McIntosh}, {McNamara}, {Megeath}, {Mochejska},
  {Munoz}, {Muzerolle}, {Naranjo}, {Narayan}, {Pahre}, {Peters}, {Peterson},
  {Rines}, {Ripman}, {Roussanova}, {Schild}, {Sicilia-Aguilar}, {Sokoloski},
  {Smalley}, {Smith}, {Spahr}, {Stanek}, {Barmby}, {Blondin}, {Stubbs},
  {Szentgyorgyi}, {Torres}, {Vaz}, {Vikhlinin}, {Wang}, {Westover}, {Woods}, \&
  {Zhao}}]{hicken09a}
{Hicken} M. {et~al.}, 2009{\natexlab{a}}, ApJ, 700, 331

\bibitem[{{Hicken} {et~al}\mbox{.}(2009{\natexlab{b}}){Hicken}, {Wood-Vasey},
  {Blondin}, {Challis}, {Jha}, {Kelly}, {Rest}, \& {Kirshner}}]{hicken09b}
{Hicken} M., {Wood-Vasey} W.~M., {Blondin} S., {Challis} P., {Jha} S., {Kelly}
  P.~L., {Rest} A., {Kirshner} R.~P., 2009{\natexlab{b}}, ApJ, 700, 1097

\bibitem[{{Hils} {et~al}\mbox{.}(1990){Hils}, {Bender}, \& {Webbink}}]{hils90}
{Hils} D., {Bender} P.~L., {Webbink} R.~F., 1990, ApJ, 360, 75

\bibitem[{{Hix} {et~al}\mbox{.}(1998){Hix}, {Khokhlov}, {Wheeler}, \&
  {Thielemann}}]{hix98}
{Hix} W.~R., {Khokhlov} A.~M., {Wheeler} J.~C., {Thielemann} F.-K., 1998, ApJ,
  503, 332

\bibitem[{{Horesh} {et~al}\mbox{.}(2011){Horesh}, {Kulkarni}, {Fox},
  {Carpenter}, {Kasliwal}, {Ofek}, {Quimby}, {Gal-Yam}, {Cenko}, {de Bruyn},
  {Kamble}, {Wijers}, {van der Horst}, {Kouveliotou}, {Podsiadlowski},
  {Sullivan}, {Maguire}, {Nugent}, {Gehrels}, {Law}, {Poznanski}, \&
  {Shara}}]{horesh11}
{Horesh} A. {et~al.}, 2011, ArXiv e-prints

\bibitem[{{Howell}(2011)}]{howell11}
{Howell} D.~A., 2011, Nature Communications, 2

\bibitem[{{Howell} {et~al}\mbox{.}(2006){Howell}, {Sullivan}, {Nugent},
  {Ellis}, {Conley}, {Le Borgne}, {Carlberg}, {Guy}, {Balam}, {Basa},
  {Fouchez}, {Hook}, {Hsiao}, {Neill}, {Pain}, {Perrett}, \&
  {Pritchet}}]{howell06}
{Howell} D.~A. {et~al.}, 2006, Nature, 443, 308

\bibitem[{{Iben} \& {Tutukov}(1984)}]{iben84}
{Iben}, Jr. I., {Tutukov} A.~V., 1984, ApJS, 54, 335

\bibitem[{{Iben} \& {Tutukov}(1985)}]{iben85}
{Iben}, Jr. I., {Tutukov} A.~V., 1985, ApJS, 58, 661

\bibitem[{{Iben} {et~al}\mbox{.}(1998){Iben}, {Tutukov}, \&
  {Fedorova}}]{iben98}
{Iben}, Jr. I., {Tutukov} A.~V., {Fedorova} A.~V., 1998, ApJ, 503, 344

\bibitem[{{Jordan} {et~al}\mbox{.}(2008){Jordan}, {Fisher}, {Townsley},
  {Calder}, {Graziani}, {Asida}, {Lamb}, \& {Truran}}]{jordan08}
{Jordan}, IV G.~C., {Fisher} R.~T., {Townsley} D.~M., {Calder} A.~C.,
  {Graziani} C., {Asida} S., {Lamb} D.~Q., {Truran} J.~W., 2008, ApJ, 681, 1448

\bibitem[{{Leaman} {et~al}\mbox{.}(2011){Leaman}, {Li}, {Chornock}, \&
  {Filippenko}}]{leaman11}
{Leaman} J., {Li} W., {Chornock} R., {Filippenko} A.~V., 2011, MNRAS, 412, 1419

\bibitem[{{Lee} {et~al}\mbox{.}(2010){Lee}, {Ramirez-Ruiz}, \& {van de
  Ven}}]{lee10}
{Lee} W.~H., {Ramirez-Ruiz} E., {van de Ven} G., 2010, ApJ, 720, 953

\bibitem[{{Li} {et~al}\mbox{.}(2011){Li}, {Bloom}, {Podsiadlowski}, {Miller},
  {Cenko}, {Jha}, {Sullivan}, {Howell}, {Nugent}, {Butler}, {Ofek}, {Kasliwal},
  {Richards}, {Stockton}, {Shih}, {Bildsten}, {Shara}, {Bibby}, {Filippenko},
  {Ganeshalingam}, {Silverman}, {Kulkarni}, {Law}, {Poznanski}, {Quimby},
  {McCully}, {Patel}, \& {Maguire}}]{li11}
{Li} W. {et~al.}, 2011, ArXiv e-prints

\bibitem[{{Lipunov} {et~al}\mbox{.}(1987){Lipunov}, {Postnov}, \&
  {Prokhorov}}]{lipunov87}
{Lipunov} V.~M., {Postnov} K.~A., {Prokhorov} M.~E., 1987, A\&A, 176, L1

\bibitem[{{Liu} {et~al}\mbox{.}(2010){Liu}, {Han}, {Zhang}, \& {Zhang}}]{liu10}
{Liu} J., {Han} Z., {Zhang} F., {Zhang} Y., 2010, ApJ, 719, 1546

\bibitem[{{Longland} {et~al}\mbox{.}(2011){Longland}, {Lor{\'e}n-Aguilar},
  {Jos{\'e}}, {Garc{\'{\i}}a-Berro}, {Althaus}, \& {Isern}}]{longland11}
{Longland} R., {Lor{\'e}n-Aguilar} P., {Jos{\'e}} J., {Garc{\'{\i}}a-Berro} E.,
  {Althaus} L.~G., {Isern} J., 2011, ApJL, 737, L34

\bibitem[{{Lor{\'e}n-Aguilar} {et~al}\mbox{.}(2009){Lor{\'e}n-Aguilar},
  {Isern}, \& {Garc{\'i}a-Berro}}]{aguilar09}
{Lor{\'e}n-Aguilar} P., {Isern} J., {Garc{\'i}a-Berro} E., 2009, A\&A, 500,
  1193

\bibitem[{{Lubow} \& {Shu}(1975)}]{lubow75}
{Lubow} S.~H., {Shu} F.~H., 1975, ApJ, 198, 383

\bibitem[{{Maoz} {et~al}\mbox{.}(2010){Maoz}, {Sharon}, \& {Gal-Yam}}]{maoz10}
{Maoz} D., {Sharon} K., {Gal-Yam} A., 2010, ApJ, 722, 1879

\bibitem[{{Marsh} {et~al}\mbox{.}(2004){Marsh}, {Nelemans}, \&
  {Steeghs}}]{marsh04}
{Marsh} T.~R., {Nelemans} G., {Steeghs} D., 2004, MNRAS, 350, 113

\bibitem[{{Marsh} \& {Steeghs}(2002)}]{marsh02}
{Marsh} T.~R., {Steeghs} D., 2002, MNRAS, 331, L7

\bibitem[{{Monaghan}(2005)}]{monaghan05}
{Monaghan} J.~J., 2005, Reports on Progress in Physics, 68, 1703

\bibitem[{{Moroni} \& {Straniero}(2009)}]{moroni09}
{Moroni} P.~G.~P., {Straniero} O., 2009, in American Institute of Physics
  Conference Series, Vol. 1111, American Institute of Physics Conference
  Series, {G.~Giobbi, A.~Tornambe, G.~Raimondo, M.~Limongi, L.~A.~Antonelli,
  N.~Menci, \& E.~Brocato}, ed., pp. 63--66

\bibitem[{{Morris} \& {Monaghan}(1997)}]{morris97}
{Morris} J.~P., {Monaghan} J.~J., 1997, Journal of Computational Physics, 136,
  41

\bibitem[{{Motl} {et~al}\mbox{.}(2007){Motl}, {Frank}, {Tohline}, \&
  {D'Souza}}]{motl07}
{Motl} P.~M., {Frank} J., {Tohline} J.~E., {D'Souza} M.~C.~R., 2007, ApJ, 670,
  1314

\bibitem[{{Nelemans}(2005)}]{nelemans05}
{Nelemans} G., 2005, in Astronomical Society of the Pacific Conference Series,
  Vol. 330, The Astrophysics of Cataclysmic Variables and Related Objects,
  {J.-M.~Hameury \& J.-P.~Lasota}, ed., p.~27

\bibitem[{{Nelemans} {et~al}\mbox{.}(2001{\natexlab{a}}){Nelemans}, {Portegies
  Zwart}, {Verbunt}, \& {Yungelson}}]{nelemans01b}
{Nelemans} G., {Portegies Zwart} S.~F., {Verbunt} F., {Yungelson} L.~R.,
  2001{\natexlab{a}}, A\&A, 368, 939

\bibitem[{{Nelemans} {et~al}\mbox{.}(2004){Nelemans}, {Yungelson}, \&
  {Portegies Zwart}}]{nelemans04}
{Nelemans} G., {Yungelson} L.~R., {Portegies Zwart} S.~F., 2004, MNRAS, 349,
  181

\bibitem[{{Nelemans} {et~al}\mbox{.}(2001{\natexlab{b}}){Nelemans},
  {Yungelson}, {Portegies Zwart}, \& {Verbunt}}]{nelemans01a}
{Nelemans} G., {Yungelson} L.~R., {Portegies Zwart} S.~F., {Verbunt} F.,
  2001{\natexlab{b}}, A\&A, 365, 491

\bibitem[{{New} \& {Tohline}(1997)}]{new97}
{New} K.~C.~B., {Tohline} J.~E., 1997, ApJ, 490, 311

\bibitem[{{Nomoto}(1982)}]{nomoto82}
{Nomoto} K., 1982, ApJ, 257, 780

\bibitem[{{Nomoto} \& {Kondo}(1991)}]{nomoto91}
{Nomoto} K., {Kondo} Y., 1991, ApJL, 367, L19

\bibitem[{{Nugent} {et~al}\mbox{.}(2011){Nugent}, {Sullivan}, {Cenko},
  {Thomas}, {Kasen}, {Howell}, {Bersier}, {Bloom}, {Kulkarni}, {Kandrashoff},
  {Filippenko}, {Silverman}, {Marcy}, {Howard}, {Isaacson}, {Maguire},
  {Suzuki}, {Tarlton}, {Pan}, {Bildsten}, {Fulton}, {Parrent}, {Sand},
  {Podsiadlowski}, {Bianco}, {Dilday}, {Graham}, {Lyman}, {James}, {Kasliwal},
  {Law}, {Quimby}, {Hook}, {Walker}, {Mazzali}, {Pian}, {Ofek}, {Gal-Yam}, \&
  {Poznanski}}]{nugent11}
{Nugent} P.~E. {et~al.}, 2011, ArXiv e-prints

\bibitem[{{Pakmor} {et~al}\mbox{.}(2011){Pakmor}, {Hachinger}, {R{\"o}pke}, \&
  {Hillebrandt}}]{pakmor11}
{Pakmor} R., {Hachinger} S., {R{\"o}pke} F.~K., {Hillebrandt} W., 2011, A\&A,
  528, A117+

\bibitem[{{Pakmor} {et~al}\mbox{.}(2010){Pakmor}, {Kromer}, {R{\"o}pke}, {Sim},
  {Ruiter}, \& {Hillebrandt}}]{pakmor10}
{Pakmor} R., {Kromer} M., {R{\"o}pke} F.~K., {Sim} S.~A., {Ruiter} A.~J.,
  {Hillebrandt} W., 2010, Nature, 463, 61

\bibitem[{{Perlmutter} {et~al}\mbox{.}(1998){Perlmutter}, {Aldering}, {della
  Valle}, {Deustua}, {Ellis}, {Fabbro}, {Fruchter}, {Goldhaber}, {Groom},
  {Hook}, {Kim}, {Kim}, {Knop}, {Lidman}, {McMahon}, {Nugent}, {Pain},
  {Panagia}, {Pennypacker}, {Ruiz-Lapuente}, {Schaefer}, \&
  {Walton}}]{perlmutter98}
{Perlmutter} S. {et~al.}, 1998, Nature, 391, 51

\bibitem[{{Piersanti} {et~al}\mbox{.}(2003{\natexlab{a}}){Piersanti},
  {Gagliardi}, {Iben}, \& {Tornamb{\'e}}}]{piersanti03b}
{Piersanti} L., {Gagliardi} S., {Iben}, Jr. I., {Tornamb{\'e}} A.,
  2003{\natexlab{a}}, ApJ, 598, 1229

\bibitem[{{Piersanti} {et~al}\mbox{.}(2003{\natexlab{b}}){Piersanti},
  {Gagliardi}, {Iben}, \& {Tornamb{\'e}}}]{piersanti03a}
{Piersanti} L., {Gagliardi} S., {Iben}, Jr. I., {Tornamb{\'e}} A.,
  2003{\natexlab{b}}, ApJ, 583, 885

\bibitem[{{Piro}(2011)}]{piro11}
{Piro} A.~L., 2011, ApJL, 740, L53

\bibitem[{{Postnov} \& {Yungelson}(2006)}]{postnov06}
{Postnov} K.~A., {Yungelson} L.~R., 2006, Living Reviews in Relativity, 9, 6

\bibitem[{{Ramsay} {et~al}\mbox{.}(2002){Ramsay}, {Hakala}, \&
  {Cropper}}]{ramsay02}
{Ramsay} G., {Hakala} P., {Cropper} M., 2002, MNRAS, 332, L7

\bibitem[{{Rappaport} {et~al}\mbox{.}(2009){Rappaport}, {Podsiadlowski}, \&
  {Horev}}]{rappaport09}
{Rappaport} S., {Podsiadlowski} P., {Horev} I., 2009, ApJ, 698, 666

\bibitem[{{Rasio} \& {Shapiro}(1995)}]{rasio95}
{Rasio} F.~A., {Shapiro} S.~L., 1995, ApJ, 438, 887

\bibitem[{{Raskin} {et~al}\mbox{.}(2011){Raskin}, {Scannapieco}, {Fryer},
  {Rockefeller\ }, \& {Timmes}}]{raskin11}
{Raskin} C., {Scannapieco} E., {Fryer} C., {Rockefeller\ } G., {Timmes} F.~X.,
  2011, ArXiv e-prints

\bibitem[{{Raskin} {et~al}\mbox{.}(2009){Raskin}, {Timmes}, {Scannapieco},
  {Diehl}, \& {Fryer}}]{raskin09}
{Raskin} C., {Timmes} F.~X., {Scannapieco} E., {Diehl} S., {Fryer} C., 2009,
  MNRAS, 399, L156

\bibitem[{{Rau} {et~al}\mbox{.}(2009){Rau}, {Kulkarni}, {Law}, {Bloom},
  {Ciardi}, {Djorgovski}, {Fox}, {Gal-Yam}, {Grillmair}, {Kasliwal}, {Nugent},
  {Ofek}, {Quimby}, {Reach}, {Shara}, {Bildsten}, {Cenko}, {Drake},
  {Filippenko}, {Helfand}, {Helou}, {Howell}, {Poznanski}, \&
  {Sullivan}}]{rau09a}
{Rau} A. {et~al.}, 2009, PASP, 121, 1334

\bibitem[{{Riess} {et~al}\mbox{.}(1998){Riess}, {Filippenko}, {Challis},
  {Clocchiatti}, {Diercks}, {Garnavich}, {Gilliland}, {Hogan}, {Jha},
  {Kirshner}, {Leibundgut}, {Phillips}, {Reiss}, {Schmidt}, {Schommer},
  {Smith}, {Spyromilio}, {Stubbs}, {Suntzeff}, \& {Tonry}}]{riess98}
{Riess} A.~G. {et~al.}, 1998, AJ, 116, 1009

\bibitem[{{R{\"o}pke} {et~al}\mbox{.}(2007){R{\"o}pke}, {Woosley}, \&
  {Hillebrandt}}]{roepke07}
{R{\"o}pke} F.~K., {Woosley} S.~E., {Hillebrandt} W., 2007, ApJ, 660, 1344

\bibitem[{{Rosswog}(2009)}]{rosswog09}
{Rosswog} S., 2009, NewAR, 53, 78

\bibitem[{{Rosswog} {et~al}\mbox{.}(2009){Rosswog}, {Kasen}, {Guillochon}, \&
  {Ramirez-Ruiz}}]{rosswog09b}
{Rosswog} S., {Kasen} D., {Guillochon} J., {Ramirez-Ruiz} E., 2009, ApJL, 705,
  L128

\bibitem[{{Rosswog} {et~al}\mbox{.}(2008){Rosswog}, {Ramirez-Ruiz}, {Hix}, \&
  {Dan}}]{rosswog08}
{Rosswog} S., {Ramirez-Ruiz} E., {Hix} W.~R., {Dan} M., 2008, Computer Physics
  Communications, 179, 184

\bibitem[{{Ruiter} {et~al}\mbox{.}(2010){Ruiter}, {Belczynski}, {Benacquista},
  {Larson}, \& {Williams}}]{ruiter10}
{Ruiter} A.~J., {Belczynski} K., {Benacquista} M., {Larson} S.~L., {Williams}
  G., 2010, ApJ, 717, 1006

\bibitem[{{Saio} \& {Nomoto}(1985)}]{saio85}
{Saio} H., {Nomoto} K., 1985, A\&A, 150, L21

\bibitem[{{Saio} \& {Nomoto}(2004)}]{saio04}
{Saio} H., {Nomoto} K., 2004, ApJ, 615, 444

\bibitem[{{Scalzo} {et~al}\mbox{.}(2010){Scalzo}, {Aldering}, {Antilogus},
  {Aragon}, {Bailey}, {Baltay}, {Bongard}, {Buton}, {Childress}, {Chotard},
  {Copin}, {Fakhouri}, {Gal-Yam}, {Gangler}, {Hoyer}, {Kasliwal}, {Loken},
  {Nugent}, {Pain}, {P{\'e}contal}, {Pereira}, {Perlmutter}, {Rabinowitz},
  {Rau}, {Rigaudier}, {Runge}, {Smadja}, {Tao}, {Thomas}, {Weaver}, \&
  {Wu}}]{scalzo10}
{Scalzo} R.~A. {et~al.}, 2010, ApJ, 713, 1073

\bibitem[{{Segretain} {et~al}\mbox{.}(1997){Segretain}, {Chabrier}, \&
  {Mochkovitch}}]{segretain97}
{Segretain} L., {Chabrier} G., {Mochkovitch} R., 1997, ApJ, 481, 355

\bibitem[{{Seitenzahl} {et~al}\mbox{.}(2009){Seitenzahl}, {Meakin}, {Townsley},
  {Lamb}, \& {Truran}}]{seitenzahl09}
{Seitenzahl} I.~R., {Meakin} C.~A., {Townsley} D.~M., {Lamb} D.~Q., {Truran}
  J.~W., 2009, ApJ, 696, 515

\bibitem[{{Shara} \& {Hurley}(2002)}]{shara02}
{Shara} M.~M., {Hurley} J.~R., 2002, ApJ, 571, 830

\bibitem[{{Shen} \& {Bildsten}(2009)}]{shen09}
{Shen} K.~J., {Bildsten} L., 2009, APJ, 699, 1365

\bibitem[{{Shen} {et~al}\mbox{.}(2011){Shen}, {Bildsten}, {Kasen}, \&
  {Quataert}}]{shen11}
{Shen} K.~J., {Bildsten} L., {Kasen} D., {Quataert} E., 2011, ArXiv e-prints

\bibitem[{{Shen} {et~al}\mbox{.}(2010){Shen}, {Kasen}, {Weinberg}, {Bildsten},
  \& {Scannapieco}}]{shen10}
{Shen} K.~J., {Kasen} D., {Weinberg} N.~N., {Bildsten} L., {Scannapieco} E.,
  2010, ApJ, 715, 767

\bibitem[{{Solheim}(2010)}]{solheim10}
{Solheim} J.-E., 2010, PASP, 122, 1133

\bibitem[{{Taam}(1980)}]{taam80}
{Taam} R.~E., 1980, ApJ, 237, 142

\bibitem[{{Taubenberger} {et~al}\mbox{.}(2011){Taubenberger}, {Benetti},
  {Childress}, {Pakmor}, {Hachinger}, {Mazzali}, {Stanishev}, {Elias-Rosa},
  {Agnoletto}, {Bufano}, {Ergon}, {Harutyunyan}, {Inserra}, {Kankare},
  {Kromer}, {Navasardyan}, {Nicolas}, {Pastorello}, {Prosperi}, {Salgado},
  {Sollerman}, {Stritzinger}, {Turatto}, {Valenti}, \&
  {Hillebrandt}}]{taubenberger11}
{Taubenberger} S. {et~al.}, 2011, MNRAS, 412, 2735

\bibitem[{{Timmes} \& {Swesty}(2000)}]{timmes00}
{Timmes} F.~X., {Swesty} F.~D., 2000, ApJS, 126, 501

\bibitem[{{Totani} {et~al}\mbox{.}(2008){Totani}, {Morokuma}, {Oda}, {Doi}, \&
  {Yasuda}}]{totani08}
{Totani} T., {Morokuma} T., {Oda} T., {Doi} M., {Yasuda} N., 2008, PASJ, 60,
  1327

\bibitem[{{Townsley} {et~al}\mbox{.}(2007){Townsley}, {Calder}, {Asida},
  {Seitenzahl}, {Peng}, {Vladimirova}, {Lamb}, \& {Truran}}]{townsley07}
{Townsley} D.~M., {Calder} A.~C., {Asida} S.~M., {Seitenzahl} I.~R., {Peng} F.,
  {Vladimirova} N., {Lamb} D.~Q., {Truran} J.~W., 2007, ApJ, 668, 1118

\bibitem[{{Warner}(1995)}]{warner95}
{Warner} B., 1995, Ap\&SS, 225, 249

\bibitem[{{Webbink}(1984)}]{webbink84}
{Webbink} R.~F., 1984, ApJ, 277, 355

\bibitem[{{Woods} {et~al}\mbox{.}(2011){Woods}, {Ivanova}, {van der Sluys}, \&
  {Chaichenets}}]{woods11}
{Woods} T.~E., {Ivanova} N., {van der Sluys} M., {Chaichenets} S., 2011, ArXiv
  e-prints

\bibitem[{{Woosley} \& {Kasen}(2011)}]{woosley11}
{Woosley} S.~E., {Kasen} D., 2011, ApJ, 734, 38

\bibitem[{{Yoon} {et~al}\mbox{.}(2007){Yoon}, {Podsiadlowski}, \&
  {Rosswog}}]{yoon07}
{Yoon} S., {Podsiadlowski} P., {Rosswog} S., 2007, MNRAS, 380, 933

\bibitem[{{Yoon} \& {Langer}(2004)}]{yoon04}
{Yoon} S.-C., {Langer} N., 2004, A\&A, 419, 645

\end{thebibliography}

\appendix

\section{Polynomial fitting functions for He detonations}
\label{sec:fitting}

Approximate formulas obtained by polynomial fitting of
the contours lines with $\tau_{\rm nuc}/\tau_{\rm dyn}=1$ are presented for the
He mass-transferring systems (see Section~\ref{sec:HeMT}). 
The contour $(\tau_{\rm nuc}/\tau_{\rm dyn})(T)=1$ shown in the upper-left
panel of Figure~\ref{fig:detonations} is approximated to better than 4\% by (in
solar units)
\begin{eqnarray}\label{eq:a1}
M_2 &=& 21.065 M_1^4 - 60.751 M_1^3 + 63.515 M_1^2\nonumber\\
&-& 28.569 M_1 + 5.1209,
\end{eqnarray}
with $M_1$ between $0.46$ and $1.13$ \msun. 

The conservative estimate for
contact He detonations $(\tau_{\rm nuc}/\tau_{\rm dyn})(\langle
T\rangle)=1$ (lower-left panel of Figure~\ref{fig:detonations}) is approximated to within $3\%$ with 
\begin{eqnarray}\label{eq:a2}
M_2 &=& 39.431 M_1^4 - 127.35 M_1^3 + 152.17 M_1^2\nonumber\\
&-& 80.163 M_1 + 16.329,
\end{eqnarray}
with $M_1$ between $0.65$ and $1.13$ \msun.

For stream detonations (see Figure~\ref{fig:streamdet}), the cubic polynomial
used to approximate the data between $0.88 \leq M_1\leq 1.2$ \Msun is:
\begin{eqnarray}\label{eq:a3}
M_2 &=& -16.318 M_1^3 + 53.568 M_1^2 - 57.809 M_1\nonumber\\
&+& 20.888. 
\end{eqnarray}
With this function we are slightly overestimating the number of systems that
may lead to a detonation for $M_1\leq 0.93$ as the curve is not single-valued in this region. Above this value
Equation (\ref{eq:a2}) approximates the data to better than $2\%$. 

\label{lastpage}

\end{document}